\RequirePackage{fix-cm}
\documentclass[smallextended,natbib]{svjour3}       
\smartqed  
\usepackage{color}
\usepackage{lineno,hyperref}
\usepackage[T1]{fontenc}
\usepackage[utf8]{inputenc}
\usepackage{listings}
\usepackage{graphicx}
\usepackage{subfigure}
\usepackage{verbatim}
\usepackage{colortbl}
\usepackage{framed}
\usepackage{url}
\usepackage{color}
\usepackage{pifont}
\newcommand{\xmark}{\ding{55}}%
\usepackage{framed}
\usepackage{tcolorbox}
\usepackage{fancybox}
\usepackage{amsmath,amssymb,amsfonts}
\usepackage{algorithmic}
\usepackage{textcomp}
\usepackage{xcolor}
\usepackage{booktabs}
\usepackage{adjustbox}
\usepackage{multirow}

\usepackage{floatrow}
\DeclareFloatFont{tiny}{\tiny}
\begin{document}

\title{Assessing Exception Handling Testing Practices in Open-Source Libraries}

\titlerunning{Assessing Exception Handling Testing Practices in Open-Source Libraries}        

\author{Luan P. Lima    \and
        Lincoln S. Rocha  \and
        Carla I. M. Bezerra \and
        Matheus Paixao
}


\institute{Luan P. Lima \at
              Federal University of Cear\'{a}, Fortaleza-CE, Brazil\\
              \email{luanpereira@great.ufc.br} 
            \and
              Lincoln S. Rocha (corresponding author) \at
              Federal University of Cear\'{a}, Fortaleza-CE, Brazil\\
              \email{lincoln@dc.ufc.br}
            \and
              Carla I. M. Bezerra \at
              Federal University of Cear\'{a}, Quixad\'{a}-CE, Brazil\\
              \email{carlailane@ufc.br}
            \and
              Matheus Paixao \at
              State University of Ceará, Fortaleza-CE, Brazil\\
              \email{matheus.paixao@uece.br}
}

\date{Received: date / Accepted: date}

\maketitle

\begin{abstract}
Modern programming languages (e.g., Java and C\#) provide features to separate error-handling code from regular code, seeking to enhance software comprehensibility and maintainability. Nevertheless, the way exception handling (EH) code is structured in such languages may lead to multiple, different, and complex control flows, which may affect the software testability. Previous studies have reported that EH code is typically neglected, not well tested, and its misuse can lead to reliability degradation and catastrophic failures. However, little is known about the relationship between testing practices and EH testing effectiveness. In this exploratory study, we (i) measured the adequacy degree of EH testing concerning code coverage (instruction, branch, and method) criteria; and (ii) evaluated the effectiveness of the EH testing by measuring its capability to detect artificially injected faults (i.e., mutants) using $7$ EH mutation operators. Our study was performed using test suites of $27$ long-lived Java libraries from open-source ecosystems. Our results show that instructions and branches within \texttt{catch} blocks and \texttt{throw} instructions are less covered, with statistical significance, than the overall instructions and branches.  Nevertheless, most of the studied libraries presented test suites capable of detecting more than $70\%$ of the injected faults. From a total of $12,331$ mutants created in this study, the test suites were able to detect $68\%$ of them.
\end{abstract}

\section{Introduction}\label{intro}

Exception handling (EH) is a forward-error recovery technique used to improve software robustness~\citep{Shahrokni2013}. An exception models an abnormal situation - detected at run time - that disrupts the normal control flow of a program~\citep{Garcia2001}. When this happens, the EH mechanism deviates the normal control flow to the abnormal (exceptional) control flow to deal with such situation. 
Mainstream programming languages (e.g., Java, Python, and C\#) provide built-in facilities to structure the exceptional control flow using proper constructs to specify, in the source code, where exceptions can be raised, propagated, and properly handled~\citep{Cacho2014a}.

Recent studies have investigated the relationship between EH code and software maintainability~\citep{Cacho2014}, evolvability~\citep{Osman2017}, architectural erosion~\citep{Filho2017}, robustness~\citep{Cacho2014a}, bug appearance~\citep{Ebert2015}, and defect-proneness~\citep{Sawadpong2016}. 
Such studies have shown that the effectiveness of EH code is directly linked to the overall software quality~\citep{dePadua2017,dePadua2018}.
To ensure and assess the EH code, developers make use of software testing, which, in this context, is referred to as EH testing~\citep{Sinha2000,Martins2014,Zhang2014}.

Despite the importance and the existence of usage patterns and guidelines for EH implementation~\citep{WirfsBrock2006,Bloch2008,Gallardo2014,jenkov2013javaeh}, this is a commonly neglected activity by developers (mostly by novice ones)~\citep{Shah2010}. 
Moreover, EH code is claimed as the least understood, documented, and tested part of a software system~\citep{Shah2008,Shah2009,Shah2010,Rashkovits2012,Kechagia2014,chang2016review,dalton2020ehtest}. 
In addition, \cite{Ebert2015} have found, in a survey with developers, that about 70\% of the software companies do not test and have no specific testing technique for EH code. 
This is a worrisome finding given the importance of EH testing effectiveness.

In the current landscape of software development, researchers commonly study open-source systems to acquire insights on many aspects of software development and quality, including architectural practices~\citep{Paixao2017}, refactoring~\citep{Bavota2015}, evolution~\citep{Koch2007} and bug fixing~\citep{Vieira2019}, to mention a few.

However, to the best of our knowledge, there is no empirical study that observes and evaluates EH testing practices in open-source software.
As a result, the software engineering community lacks a thorough and concise understanding of good and openly available EH testing practices.
This prevents the further creation of EH testing guidelines that are based on real-world software and practices instead of textbooks~\citep{hunt2003unittesting,gulati2017junit} and rules of thumb\footnote{\url{http://wiki.c2.com/?ExceptionPatterns}}.

Nevertheless, to evaluate software testing as a whole, and EH testing in specific, is not a trivial task.
First, one needs to define what constitutes a good test. 
Early in 1975, \cite{Goodenough1975} defined the concept of test criterion as a way to precisely state what constitutes a suitable software test. 
Currently, the code coverage (e.g., instruction, branch, and method) criteria have been widely used as a proxy for testing effectiveness~\citep{Ivankovic2019,Yang2019}. 
However, recent studies provide evidence that high test coverage alone is not sufficient to avoid software bugs~\citep{Antinyan2017,Kochhar2017}. 
In parallel, mutation testing (a.k.a, mutation analysis) provides a way to evaluate the effectiveness of test suites by artificially injecting bugs that are similar to real defects~\citep{Papadakis2019}.
Studies have shown that mutant detection is significantly correlated with real fault detection~\citep{Just2014}.

Hence, in this study, we report on the first empirical study that assesses and evaluates the practices of EH testing in open-source libraries.
We developed a tool, called XaviEH, to assist in these analyses.
XaviEH employs both coverage and mutation analysis as proxies for EH testing effectiveness in a certain system.
In addition, XaviEH uses tailored criteria for EH code, including EH-specific coverage measures and mutation operators.
In total, XaviEH measured the adequacy and effectiveness of EH testing of 27 long-lived Java libraries. Finally, based on the analysis by XaviEH, we assess whether there are types of EH bugs that are more difficult to detect by the studied libraries’ test suites than others.
Finally, based on the analysis by XaviEH, we ranked these libraries to assess which ones present significantly better indicators of EH testing effectiveness.

The main contributions of this paper are listed as follows:
\begin{itemize}
    \item The first empirical study to evaluate adequacy and effectiveness of EH testing practices in open-source libraries.
    \item A tool, called XaviEH, to automatically assess the adequacy and effectiveness of EH testing in a software system.
    \item A dataset concerning the analysis of 27 long-lived Java libraries regarding their EH testing practices~\citep{replication_package}.
\end{itemize}

Overall, our findings suggest that EH code is, in general, less covered than regular code (i.e., non-EH). 
Additionally, we provide evidence that the code within the \texttt{catch} blocks and \texttt{throw} statements have a low coverage degree. 
However, despite not being well-covered, the mutation analysis shows that the test suites are able to detect artificial EH-related faults. 

The remainder of this paper is organized as follows. Section~\ref{sec:background} provides a background for our study. Section~\ref{sec:settings} presents the experimental design of our study. The study results are presented in Section~\ref{sec:results}. In Section~\ref{sec:discussion}, our results and implications for researchers and practitioners are discussed. Section~\ref{sec:threats} presents the threats to validity. Section~\ref{sec:relatedwork} addresses the related work, and at last, Section~\ref{sec:conclusion} concludes the paper and points out directions for future work.
\section{Background}\label{sec:background}

In this section, we describe the general concepts and definitions that provide a background to our study.

\subsection{Software Test Criteria and Adequacy}\label{sec:TestCriteriaAdequacy}

\cite{Goodenough1975} state that a software test adequacy criterion defines ``\emph{what properties of a program must be exercised to constitute a `thorough' test, i.e., one whose successful execution implies no errors in a tested program}''. 
To guarantee the correctness of adequately tested programs, they proposed reliability and validity requirements of test criteria~\citep{Zhu1997}. 
The former requires that a test criterion always produce consistent test results (i.e., if the program is tested successfully on a certain test set that satisfies the criterion, then the program should also be tested successfully on all other test sets that satisfies the criterion). 
The later requires that the test should always produce a meaningful result concerning the program under testing (i.e., for every error in a program, there exists a test set that satisfies the criterion and it is capable of revealing the error).

Code coverage (also known as test coverage) is a metric to assess the percentage of the source code executed by a test suite.
Code coverage is commonly employed as a proxy for test adequacy~\citep{Kochhar2017}. 
The percentage of code executed by test cases can be measured according to various criteria, such as: statement coverage, branch coverage, and function/method coverage~\citep{Antinyan2017}. 
Statement coverage is the percentage of statements in a source file that have been exercised during a test run. 
Branch coverage is the percentage of decision blocks in a source file that have been exercised during a test run. 
Function/Method coverage, is the percentage of all functions/methods in a source file that have been exercised during a test run.
For the rest of this paper, we use code coverage and test coverage interchangeably.

\subsection{Mutation Testing and Analysis}
\label{subsec:mutation}

Mutation analysis is a procedure for evaluating the degree to which a program is properly tested, that is, to measure a test suite's effectiveness. According to~\cite{Offutt2016}, mutation testing is commonly used as a ``gold standard'' in experimental studies for comparative evaluation of other test criteria.
Mutation testing evaluates a certain test suite by injecting artificial defects in the source code.
In this context, a test suite that is able to identify artificial defects is likely to be able to pinpoint real defects when these occur.
Hence, to maximize mutation testing's ability to measure the effectiveness of a test suite, one must inject artificial defects that are as close as possible to real defects~\citep{Just2014,Papadakis2018}.

A version of a software system with an artificially inserted fault is called a mutant. 
Mutation operators are rule-based program transformations used to create mutants from the original source code of a software system.
When executing the test suite of a system in both the original and mutant code, if the mutant and the original code produce different outputs in at least one test case, the fault is detected, i.e., the mutant can be killed by the test suite.

Consider $M(s)$ to be the set of mutants created for system $s$ and $KM(s)$ the set of killed mutants for system $s$.
Mutation score, as detailed in Equation (\ref{eq:mutation_score}), indicates the ratio of killed mutants compared to all created mutants~\citep{Zhu1997,Offutt2016}.
Mutation score indicates the effectiveness of a certain test suite, as it evaluates the test suite's ability to find defects.

\begin{equation}
\label{eq:mutation_score}
    \mathit{MutationScore(s)} = \frac{|\mathit{KM(s)}|}{|\mathit{M(s)}|}
\end{equation}

There are cases where it is not possible to find a test case that could kill a mutant. The mutant is behaviorally equivalent to the original program. This kind of mutants are referred to as equivalent mutants. Therefore, to obtain a more accurate mutation score, it is necessary remove the equivalent mutants $E(s)$ from the set of created mutants, resulting in improved definition of mutation score, as depicted in Equation (\ref{eq:mutation_score_equivalent}).

\begin{equation}
\label{eq:mutation_score_equivalent}
    \mathit{MutationScore(s)} = \frac{|\mathit{KM(s)}|}{|\mathit{M(s)} - \mathit{E(s)}|}
\end{equation}

A certain variant of a software system is considered a first-order mutant when only a single artificial defect has been introduced.
Differently, higher-order mutants are the ones generated by combining more than one mutation operator.
We focus on first-order mutants for this study (see Section~\ref{subsec:XaviEH}).

\subsection{Java Exception Handling}

In the Java programming language, ``\emph{an exception is an event, which occurs during the execution of a program, which disrupts the normal flow of the program's instructions}''~\citep{Gallardo2014}. When an error occurs inside a method, an exception is raised. In Java, the raising of an exception is called \textit{throwing}. Exceptions are represented as objects following a class hierarchy and can be divided into two categories: checked and unchecked. Checked exceptions are all exceptions that inherit, directly or indirectly, from Java's \texttt{Exception} class, except those that inherit, directly or indirectly, from \texttt{Error} or \texttt{RuntimeException} classes, named unchecked ones. Checked exceptions represent exceptional conditions that, hypothetically, a robust software should be able to recover from. Unchecked exceptions represent an internal (\texttt{RuntimeException}) or an external (\texttt{Error}) exceptional condition that a software usually cannot anticipate or recover from. In Java, only the handling of checked exceptions is mandatory, which obligate developers to write error-handling code to catch and handle them.

When an exception is raised, the execution flow is interrupted and deviated to a specific point where the exceptional condition is handled. In Java, exceptions can be raised using the \texttt{throw} statement, signaled using the \texttt{throws} statement, and handled in the \texttt{try-catch-finally} blocks. The ``\texttt{throw new E()}'' statement is an example of \textit{throwing} the exception \texttt{E}. The ``\texttt{public void m() throws E,T}'' shows how the \texttt{throws} clause is used in the method declaration to indicate the signaling of exceptions \texttt{E} and \texttt{T} to the method that call \texttt{m()}. 

The \texttt{try} block is used to enclose the method calls that might throw an exception. If an exception occurs within the \texttt{try} block, that exception is handled by an exception handler associated with it. 
Handlers are represented by \texttt{catch} blocks that are written right below the respective \texttt{try} block. 
Multiple \texttt{catch} blocks can be associated with a \texttt{try} block. 
Each \texttt{catch} block catches a specific exception type and encloses the exception handler code. 
The \texttt{finally} block is optional, but when declared, it always executes when the \texttt{try} block finishes, with or without an exception occurring and/or being handled. 
Finally blocks are commonly used for coding cleanup actions.
\section{Experimental Design}\label{sec:settings}

Our study aims at investigating practices for EH testing in open-source libraries.
To achieve this, we selected 27 long-lived Java libraries to serve as subjects in our empirical evaluation (see Section~\ref{subsec:libraries}).
Hence, we ask the following research questions:

\bigskip
\noindent \textbf{RQ1}.~\textit{What is the test coverage of EH code in long-lived Java libraries}?

First, we measure EH testing adequacy in terms of code coverage measures.
We employ a variant of long-established coverage criteria in the literature (see Section~\ref{sec:CodeCoverage}) to provide the first insight regarding the extent to which the test suites of the studied Java libraries exercise EH code.

\bigskip
\noindent \textbf{RQ2}.~\textit{What is the difference between EH and non-EH code coverage in long-lived Java libraries}?

In addition to measuring the coverage of EH code, we also measure the coverage of non-EH code in each library.
By controlling the EH coverage with its non-EH counterpart, we can reason on how EH testing differs from other testing activities to better understand how (in)adequate EH testing may be.

\bigskip
\noindent \textbf{RQ3}.~\textit{What is the effectiveness of EH testing in long-lived Java libraries}?

We employ mutation testing to assess the effectiveness of EH testing.
By leveraging EH mutation operators derived from real EH bugs, we create artificial defects that are similar to EH bugs found in real-world software libraries.
Next, we measure the mutation score and use it as a proxy for the effectiveness of EH testing.

\bigskip
\noindent \textbf{RQ4}.~\textit{To what extent are there EH bugs that are statistically harder to detect by test suites of long-lived Java libraries?}

We employ a combination of the Friedman~\citep{Friedman1940} and Nemenyi \citep{Demsar2006} tests to statistically assess whether there are types of EH bugs that are more difficult to detect by the studied libraries' test suites than others. 
We aspire to find a set of EH bugs that developers must be aware of during testing, aiming at fostering knowledge and improving the effectiveness of EH testing.

The rest of this section details the methodology employed in our empirical study to answer the research questions presented above.
The complete dataset, source code and results for this empirical study are available in our replication package~\citep{replication_package}.

\subsection{Selection of Long-lived Java Libraries}
\label{subsec:libraries}

Our study focuses on the study of EH testing. 
However, EH is not a trivial activity in software development~\citep{Shah2010}.
First, the need for EH commonly arises as systems evolve and are exposed to a wide range of usage scenarios that expose runtime flaws~\citep{Cacho2014, dePadua2018, Chen2019}.
Second, EH testing is considered more challenging than non-EH testing due to its (i) complex runtime and (ii) \textit{flakiness}~\citep{Zhang2014,Eck2019}.
Flaky tests are the ones that can intermittently pass or fail even for the same code version~\citep{luo2014empirical}.
Consider a test aimed at reproducing the (un)availability of resources at runtime, such as internet connection or databases, for example.
Most of such resources cannot be easily mocked, and the test execution is bound to an external state of the system, which may cause a flaky behavior.

Hence, to properly study EH testing, we need long-lived subject systems that cater to a large number of users and usage scenarios. In addition, we need systems with reputably good quality to maximize the chances that the development team is versed and employ good practices in both EH and testing.

Therefore, we turned our attention to the Apache Software Foundation ecosystem.\footnote{\url{https://apache.org/index.html\#projects-list}}
The Apache Foundation is a well-known open-source software community that leads the continuous development of open-source general-purpose software solutions.
Not only this community hosts long-lived systems in active development (Apache's Commons Collections library, for instance, is now 17 years old) but it is also known to follow good software engineering practices, where its systems have been the object of a plethora of previous empirical studies~\citep{Shi2011,Barbosa2014,Ahmed2016,Schwartz2018,Hilton2018,Digkas2018,Vieira2019,Zhong2019}.

For this particular study, we considered libraries of the Apache Commons Project, which is an Apache project focused on all aspects of reusable Java components.\footnote{\url{https://commons.apache.org/}} We focused on libraries because they tend to be more generic and present more usage scenarios than other systems. 
As a selection criteria for our study, a library should: (i) be developed in Java; (ii) employ Maven or Gradle as build system; (iii) present an automatically executable and passing test suite; (iv) be a long-lived system; and (v) be correctly handled by Spoon~\citep{Pawlak2016}, one of the tools we used to build XaviEH (see Section~\ref{subsec:XaviEH}).
To identify long-lived libraries, we computed the distribution of the age of all Apache Commons's libraries in years.
Hence, we considered long-lived systems to be all libraries above the 3rd quartile in the distribution, which, for this study, represent systems with more than 11 years of active development.

As a result, we selected 21 libraries out of the 96 available in Apache Commons. 
We provide details about each selected library in the first section of Table~\ref{tab:libraries}. 
Nevertheless, while fit for our empirical study, to consider only libraries from the Apache community would represent a threat to the study's generability and diversity~\citep{Nagappan2013}. Hence, we selected 6 additional non-Apache libraries that adhere to the same inclusion criteria discussed above. These were selected considering their ranking on open-source platforms, such as GitHub, and personal experience from the authors in using these libraries. The additional libraries are depicted in the second section of Table~\ref{tab:libraries}. In total, our empirical study considered 27 long-lived libraries from different open-source ecosystems.

\begin{table}
\renewcommand{\arraystretch}{1.2}
\caption{Summary of selected libraries. While the first section depicts the libraries from Apache Commons, the second section indicates the selected libraries from other ecosystems. We provide the version we studied of each library followed by size metrics, such as LoC, number of \texttt{throw} instructions, number of \texttt{try} blocks etc.}
\centering
\label{tab:libraries}
\footnotesize
\resizebox{\columnwidth}{!}{
\begin{tabular}{l|r|r|r|r|r|r|r|r}
    \toprule
    \multicolumn{1}{c}{\textbf{Library}} &  \multicolumn{1}{|c}{\textbf{Version}} & \multicolumn{1}{|c}{\textbf{\#LoC}} &  \multicolumn{1}{|c}{\textbf{\#Classes}} & \multicolumn{1}{|c}{\textbf{\#Throw}} & \multicolumn{1}{|c}{\textbf{\#Try}} & \multicolumn{1}{|c}{\textbf{\#Catch}} & \multicolumn{1}{|c}{\textbf{\#Finally}} & \multicolumn{1}{|c}{\textbf{\#Years}}\\
    \midrule
    
    \textbf{BCEL} & 6.2 & 61100 & 344 & 406 & 147 & 143 & 5 & 18\\
    
    \textbf{BeanUtils} & 1.9.3 & 32150 & 98 & 364 & 126 & 164 & 0 & 18\\

    \textbf{CLI} & 1.4 & 6245 & 21 & 29 & 12 & 11 & 1 & 17\\
    
    \textbf{Codec} & 1.11 & 18559 & 55 & 97 & 28 & 22 & 8 & 16\\
    
    \textbf{Collections} & 4.2 & 68319 & 270 & 725 & 28 & 44 & 1 & 18\\
    
    \textbf{Compress} & 1.18 & 47741 & 183 & 425 & 137 & 73 & 39 & 16\\
    
    \textbf{Configuration} & 2.4 & 66869 & 178 & 306 & 235 & 159 & 96 & 16\\
    
    \textbf{DBCP} & 2.5 & 23132 & 50 & 279 & 796 & 846 & 23 & 18\\
    
    \textbf{DbUtils} & 1.7 & 8850 & 39 & 46 & 41 & 40 & 20 & 16\\
    
    \textbf{Digester} & 3.3.2 & 22858 & 132 & 110 & 68 & 72 & 7 & 18\\
    
    \textbf{Email} & 1.5 & 6115 & 19 & 74 & 36 & 32 & 9 & 15\\
    
    \textbf{Exec} & 1.3 & 4600 & 26 & 29 & 23 & 23 & 6 & 14\\
    
    \textbf{FileUpload} & 1.3.3 & 6884 & 23 & 50 & 25 & 26 & 6 & 17\\
    
    \textbf{Functor} & 1.0 & 17617 & 135 & 115 & 0 & 0 & 0 & 16\\
    
    \textbf{IO} & 2.6 & 28691 & 112 & 292 & 106 & 80 & 8 & 17\\
    
    \textbf{Lang} & 3.8.1 & 78174 & 124 & 380 & 76 & 81 & 5 & 17\\
    
    \textbf{Math} & 3.6.1 & 223110 & 740 & 1494 & 118 & 124 & 4 & 16\\
    
    \textbf{Net} & 3.6 & 47107 & 175 & 159 & 174 & 180 & 24 & 17\\
    
    \textbf{Pool} & 2.6.1 & 13629 & 33 & 79 & 132 & 70 & 79 & 18\\
    
    \textbf{Proxy} & 1.0 & 4112 & 37 & 36 & 23 & 31 & 0 & 11\\
    
    \textbf{Validator} & 1.6 & 17677 & 62 & 68 & 40 & 49 & 1 & 17\\
    \midrule
    
    \textbf{Gson} & 2.8.5 & 14863 & 52 & 222 & 56 & 75 & 5 & 11\\
    
    \textbf{Hamcrest} & 2.1 & 7834 & 77 & 19 & 12 & 13 & 0 & 13\\
    
    \textbf{Jsoup} & 1.11.3 & 18111 & 55 & 46 & 33 & 32 & 3 & 11\\
    
    \textbf{JUnit} & 4.12 & 17200 & 149 & 101 & 99 & 119 & 17 & 19\\
    
    \textbf{Mockito} & 2.23.11 & 33505 & 297 & 236 & 91 & 98 & 20 & 12\\
    
    \textbf{X-Stream} & 1.4.11.1 & 37475 & 313 & 461 & 248 & 362 & 19 & 16\\    
    \bottomrule
\end{tabular}
}
\end{table}

After selecting the 27 libraries employed in the study, we performed the data collection. 
On March 2019, we downloaded the latest available release of each library in which we could automatically build and execute the test suite without any failing test. 

\subsection{Assessing Exception Handling Testing with XaviEH}
\label{subsec:XaviEH}

To perform our study, we developed the XaviEH tool. Given a certain software system, XaviEH is able to automatically perform an analysis regarding the system's practices on EH testing. It provides a report on the adequacy and effectiveness of the system's test suite when testing EH code. However, before explaining in detail the XaviEH execution steps, it is worth to mention the limitations of existing tools for test coverage regarding EH code. 

\begin{figure}
	\centering
	\includegraphics[scale=0.5]{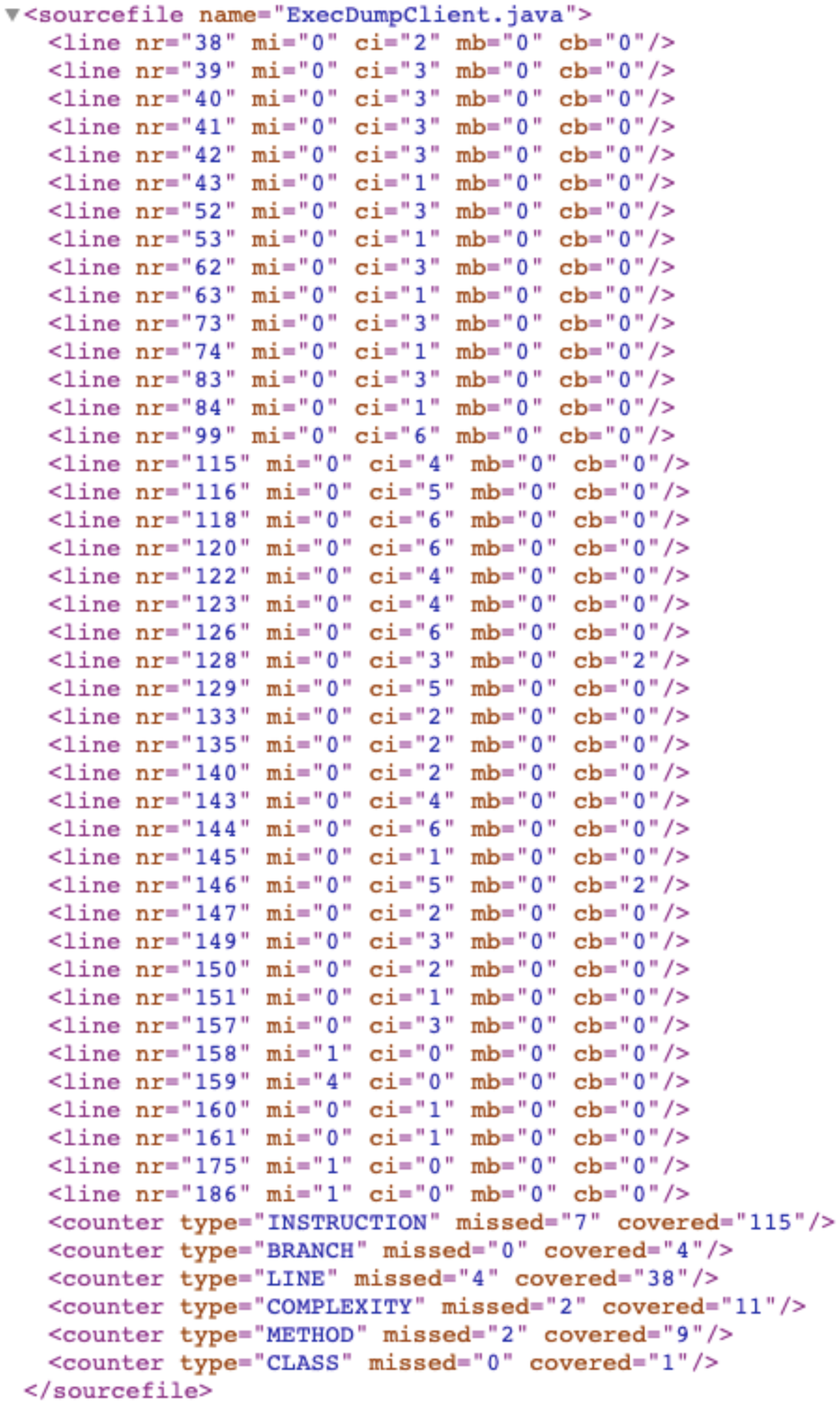}
	\caption{Fragment of a JaCoCo XML-based test coverage report.}
	\label{fig:JaCoCoReport}
\end{figure}

Overall, existing test coverage tools (e.g., Cobertura\footnote{\url{http://cobertura.github.io/cobertura/}}, JaCoCo\footnote{\url{https://www.jacoco.org/jacoco/}}, and OpenClover\footnote{\url{http://openclover.org/}}) compute all instruction, branch, and method coverages, and outputs such information within a coverage report following a specific format. In JaCoCo, for instance, one can choose to generate an XML-based report, which follows a well-defined DTD format\footnote{\url{https://www.jacoco.org/jacoco/trunk/coverage/report.dtd}}. Fig.~\ref{fig:JaCoCoReport} shows an example fragment of a JaCoCo XML-based report\footnote{\url{https://www.jacoco.org/jacoco/trunk/coverage/jacoco.xml}}. On the upper part of Fig.~\ref{fig:JaCoCoReport}, one can see the class name that was the target of the test coverage. In the middle of Fig.~\ref{fig:JaCoCoReport}, for each line of code (comments and empty lines are not taken into account), the report gives the following information: the line number in the source code file (nr), the number of missed instructions (mi), covered instructions (ci), missed branches (mb), and covered branches (cb). Finally, at the bottom of Fig.~\ref{fig:JaCoCoReport}, the report provides a summary for the class under consideration concerning the JaCoCo general coverage metrics (e.g., instruction, branch, line of code, and method). From this report, one may see that the main limitation of JaCoCo (which is shared by other test coverage tools) is that it does not differ EH code from non-EH code. Thus, this information remains hidden in the coverage report. To overcome this limitation, XaviEH employs static code analysis to determine which parts of the coverage report are related to EH code and non-EH code.

\begin{figure}
	\centering
	\includegraphics[scale=0.3]{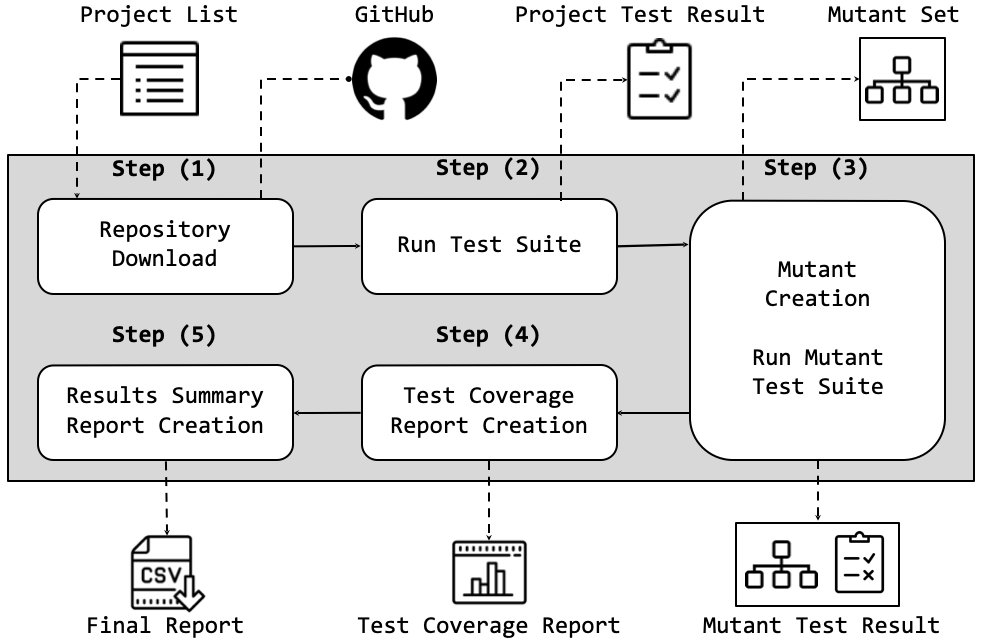}
	\caption{Internal steps performed by XaviEH when evaluating the EH testing practices of a software system.}
	\label{fig:steps}
\end{figure}

Fig.~\ref{fig:steps} illustrates the execution steps of XaviEH for a software system.
In Step (1), XaviEH obtains the system's source code. 
In case the system is hosted on GitHub, one can provide the GitHub URL, and XaviEH will use JGit\footnote{\url{https://www.eclipse.org/jgit/}} to download the source code.
Otherwise, one can simply provide the local source code path to XaviEH.

In Step (2), XaviEH executes the system's test suite to verify that all tests are passing. 
This is necessary to ensure that the next steps will be correctly executed.
XaviEH uses the \texttt{maven-invoker}\footnote{\url{https://maven.apache.org/shared/maven-invoker/}} and \texttt{gradle-tooling-api}\footnote{\url{https://docs.gradle.org/current/userguide/embedding.html}} libraries to run the tests automatically and to identify whether a test suite is passing or not.

Step (3) involves the mutation analysis. 
XaviEH generates all possible first-order mutants of the system being analyzed. 
To do this, XaviEH first searchs the system's source code to identify all classes eligible for mutation. 
In this study, a class is considered eligible if it has any code structure that can be affected (mutated) by at least one of the seven mutation operators we employ (see Section~\ref{sec:MutationOperators}). 
By doing so, XaviEH creates an in-memory data structure that tracks eligible classes and mutation operators that can be applied to each class. 

Next, for each eligible class, XaviEH applies all mutation operators that can be applied to the class, recording which operator was applied to which class. 
A mutation operator may be applied to an eligible class more than once. 
In this case, XaviEH ensures that successive changes made by a specific mutation operator within the same eligible class do not affect the same location twice, preventing duplicate mutants from being generated. 
To perform both code search and mutant generation tasks, XaviEH uses Spoon\footnote {\url{https://spoon.gforge.inria.fr/}}, a library for parsing and transforming Java source code.

For each mutant, XaviEH runs the system's test suite against it, recording the passing and failing test cases. 
This task is also performed using the \texttt{maven-invoker} and \texttt{gradle-tooling-api}.
Finally, at the end of Step~(3), XaviEH provides a mutation analysis report of the system being analyzed.

In Step (4), XaviEH performs the test coverage analysis using JaCoCo, a Java code coverage library for monitoring and tracking code coverage. 
JaCoCo was chosen for this step for a couple of reasons.
First, some Apache projects already employ JaCoCo as their official tool for test code analysis within their projects\footnote{\url{https://commons.apache.org/proper/commons-io/project-reports.html}}.
Second, JaCoCo is the tool of choice in previous related empirical studies~\citep{Saha2018,Turner2016}.
Hence, by employing a tool used by both practitioners and researchers, we would enhance XaviEH's relevance and actionability.
Finally, one of the authors of the paper had previous experience with JaCoCo, which gave us more control over its integration into XaviEH.

Hence, XaviEH uses JaCoCo to generate a XML-based coverage report for each system under analysis. Next, for each system under consideration, XaviEH uses information from the coverage report and the static code analysis data provided by Spoon to determine what code parts referred to in the JaCoCo report are related or not to EH code. For instance, XaviEH employs Spoon to identify which lines of code of a class are within \texttt{try}-\texttt{catch}-\texttt{finally} blocks and triangulates such information with code line numbers provided by JaCoCo report to track what instructions and branches within  \texttt{try}-\texttt{catch}-\texttt{finally} blocks are covered or not. Finally, for each system under analysis, XaviEH collects all information needed to compute a suite of 24 test coverage metrics, as detailed in Section~\ref{sec:CodeCoverage}.

Finally, in Step~(5), XaviEH summarizes the mutant and coverage analysis for the system under study. It generates two main reports in CSV files. The first one contains a pair of values (in the columns) needed to compute the  coverage metrics. For instance, consider the metric \texttt{CATCH\_IC} (see Section~\ref{sec:CodeCoverage}). In the report, we have the number of instructions missed (\texttt{CATCH\_MI}) and covered (\texttt{CATCH\_CI}) within the \texttt{catch} blocks. In this case, \texttt{CATCH\_IC} is computed as \texttt{CATCH\_CI}/(\texttt{CATCH\_MI} + \texttt{CATCH\_CI}). The second report file contains, for each mutation operator, the number of mutants killed and alive (in the columns), making it easy to compute the mutation score (see Section~\ref{subsec:mutation}). It is important to notice that XaviEH does not perform any kind of statistical analysis and, as a limitation, it only can be employed in the analysis of programs written in Java, automatically built using Maven or Gradle, and that use JUnit to run its unit tests.

\subsection{Mutation Operators and Analysis}
\label{sec:MutationOperators}

As discussed in Section~\ref{subsec:mutation}, we used mutation testing to assess the effectiveness of the test suites under study on identifying defects related to EH code.
Hence, we employed a set of EH-specific mutation operators proposed in previous studies~\citep{ji2009new,kumar2011new}.
Such mutation operators are based on real-world defects collected from empirical studies in open-source software.
Thus, they mirror real defects introduced by developers.
In total, we employed 7 mutation operators, as detailed in Table~\ref{tab:EHMutationOperators}.
The first 5 mutation operators (\texttt{CBR}, \texttt{CBI}, \texttt{CBD}, \texttt{PTL}, and \texttt{CRE}) were proposed by~\cite{ji2009new}, and the final 2 operators (\texttt{FBD} and \texttt{TSD}) were proposed by~\cite{kumar2011new}.

\begin{table}
\renewcommand{\arraystretch}{1.2}
\caption{Mutation operators employed in this study. All operators are based on real-world defects from open-source systems.}
\centering
\label{tab:EHMutationOperators}
\footnotesize
\resizebox{\columnwidth}{!}{
\begin{tabular}{c|p{0.77\columnwidth}}
    \toprule
    \multicolumn{1}{c}{\textbf{Operator}} & \multicolumn{1}{|c}{\textbf{Transformation in the Code}}\\
    \midrule

     \texttt{CBR} & \textit{Catch Block Replacement}. Replaces the \texttt{catch} block with exception types \textcolor{blue}{present in the} invoking exception hierarchy (\textcolor{blue}{IEH})~\citep{ji2009new}.\\
    \midrule
     \texttt{CBI} & \textit{Catch Block Insertion}. Creates complete \texttt{catch} modules to conceal all types of exceptions~\citep{ji2009new}.\\
    \midrule
     \texttt{CBD} & \textit{Catch Block Deletion}. Deletes the whole \texttt{catch} block to propagate the thrown exceptions~\citep{ji2009new}.\\
    \midrule
     \texttt{PTL} & \textit{Placing Try Block Later}. Brings into the \texttt{try} block, statements placed after the \texttt{try} block that reference variables inside the \texttt{try} block~\citep{ji2009new}.\\
    \midrule
     \texttt{CRE} & \textit{Catch and Rethrow Exception}. Re-throws the caught exceptions which are propagated to the upper modules~\citep{ji2009new}.\\
    \midrule
     \texttt{FBD} & \textit{Finally Block Deletion}. Deletes the whole \texttt{finally} block to propagate the thrown exceptions~\citep{kumar2011new}.\\
    \midrule
    \texttt{TSD} & \textit{Throw Statement Deletion}. Deletes the \texttt{throw} statement that should raise an exception~\citep{kumar2011new}.\\
    \bottomrule
\end{tabular}
}
\end{table}

\begin{figure}
	\centering
	\includegraphics[scale=0.65]{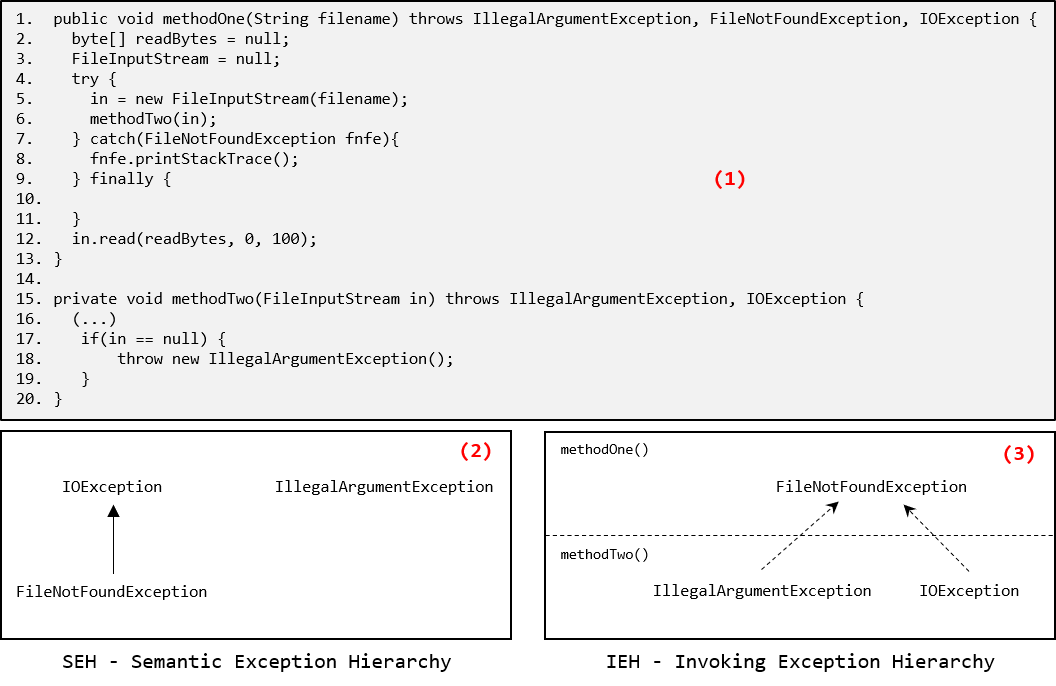}
	\caption{An illustrative example of exception handling source code (1), semantic exception hierarchy (2), and invoking exception hierarchy (3) adapted from~\citep{ji2009new}.}
	\label{fig:EHCodeExample}
\end{figure}

The meaning of some mutation operators are very straightforward such as \texttt{CBD}, \texttt{CRE}, \texttt{FBD}, and \texttt{TSD} but the others are not so simple. Thus, to ease the understanding, we provide in Fig.~\ref{fig:EHCodeExample} an illustrative exception handling code sample (1) and its two exception hierarchies (2) and (3). The semantic exception hierarchy (2) indicates the inheritance relationship between the exception types involved in the EH scenario. \textcolor{blue}{The invoking exception hierarchy (3) aims at organizing the structure of the program according to the relationship of different exception handlers using information from the method calls' chain. In (3), the exception type of the \texttt{catch} block attached to the \texttt{try} block in \texttt{methodOne()} (caller method) represents the root node (i.e., \texttt{FileNotFoundException}). Each type of exception (i.e., \texttt{IllegalArgumentException} and \texttt{IOException}) thrown by \texttt{methodTwo()}, which is called in the \texttt{try} block of \texttt{methodOne()}, is linked to the root node.} These hierarchies help mutation operators determine how to transform the original program and inject defects.

\begin{figure}
	\centering
	\includegraphics[scale=0.7]{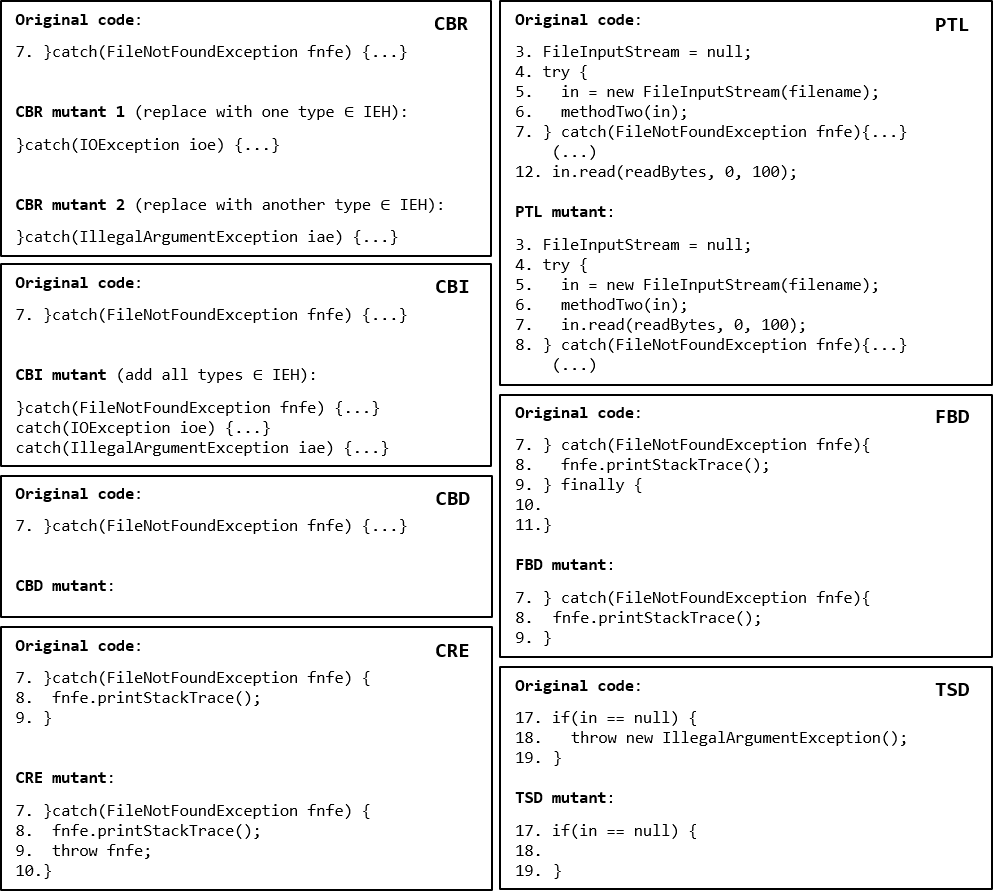}
	\caption{An illustrative example of the transformations performed by each EH mutation operator regarding the EH code example shown in Fig.~\ref{fig:EHCodeExample}.}
	\label{fig:EHOperatorsInAction}
\end{figure}

Based on the EH code of Fig.~\ref{fig:EHCodeExample}, we provide in Fig.~\ref{fig:EHOperatorsInAction} an illustrative example of the transformations performed by each EH mutation operator.

\subsection{Code Coverage Metrics}
\label{sec:CodeCoverage}

In this study, we adopted three different criteria to measure code coverage: instruction, branch, and method coverage.
We have used the XaviEH tool to compute the code coverage metrics. 
Instead of statements, XaviEH (through JaCoCo) computes the code coverage by analyzing bytecode instructions. 
Thus, we chose instruction coverage instead of statement coverage for compliance purposes.

We considered four sets of coverage metrics.
In the first set, we computed the overall instruction, branch and method coverage, i.e., considering the library's entire code base (EH code \textbf{and} non-EH code).
This is necessary for us to have a baseline of each library's general coverage, so that we can assess whether the EH code coverage presents any disparity when compared to the overall coverage.
We detail the overall code coverage metrics in the first section of Table~\ref{tab:GeneralCodeCoverage}.
Next, the \texttt{TRY\_CATCH\_BC} coverage metric represents a specific EH-related metric that counts different \texttt{catch} blocks associated with a \texttt{try} block being selected depending on the raised exception type as a case of branching. It is important to notice that \texttt{BC} metric of the first section of Table~\ref{tab:GeneralCodeCoverage} does not take \texttt{try-catch} statements into account to compute branch coverage, which make \texttt{TRY\_CATCH\_BC} and \texttt{BC} distinct coverage metrics.

Next, the coverage metrics in the third set are tailored for EH code.
It considers the instructions and branches inside \texttt{try}, \texttt{catch} and \texttt{finally} blocks.
These are detailed in the third section of Table~\ref{tab:GeneralCodeCoverage}.
Finally, the fourth set of coverage metrics is applied to the code outside \texttt{try}, \texttt{catch} and \texttt{finally} blocks, i.e., non-EH code.
Considering the overall coverage metrics as the baseline (first section), the non-EH coverage metrics (fourth section) are mathematical complements to the EH coverage metrics (third section).
For instance, \texttt{BC = EH\_BC + NON\_EH\_BC}.

\begin{table}
\renewcommand{\arraystretch}{1.2}
\caption{Code coverage metrics computed by XaviEH. The first and second set of metrics are computed considering the library's entire code base. The third set of coverage metrics are specific for exception handling code. Finally, the fourth set is composed of mathematical complements to the exception handling coverage metrics in the third section.}
\centering
\label{tab:GeneralCodeCoverage}
\footnotesize
\resizebox{\columnwidth}{!}{
\begin{tabular}{l|p{0.77\columnwidth}}
    \toprule
    \multicolumn{1}{c}{\textbf{Metric}} & \multicolumn{1}{|c}{\textbf{Meaning}}\\
    \midrule
    
    \texttt{IC} & \textit{Instruction Coverage}. The percentage of instructions exercised by the test suite.\\
    \midrule
    
    \texttt{BC} & \textit{Branch Coverage}. The percentage of branches exercised by the test suite.\\
    \midrule
    
    \texttt{MC} & \textit{Method Coverage}. The percentage of methods exercised by the test suite.\\
    \midrule
    \midrule

    \texttt{TRY\_CATCH\_BC} & \textit{Try-Catch Branch Coverage}. The percentage of branches of \texttt{try-catch} exercised by the test suite. Each \texttt{catch} block associated with a \texttt{try} block can be seen as a possible branch based on the exception type.\\
    \midrule
    \midrule
    
    \texttt{EH\_IC} & \textit{Exception Handling Instruction Coverage}. Instructions, \texttt{catch}, and \texttt{finally} blocks plus all \texttt{throw} instructions exercised by the test suite.\\
    \midrule
    
    \texttt{EH\_BC} & \textit{Exception Handling Branch Coverage}. The percentage of branches in \texttt{try}, \texttt{catch}, and \texttt{finally} blocks exercised by the test suite.\\
    \midrule

    \texttt{TRY\_IC} & \textit{Try Instruction Coverage}. The percentage of instructions in \texttt{try} blocks exercised by the test suite.\\
    \midrule
    
    \texttt{TRY\_BC} & \textit{Try Branch Coverage}. The percentage of branches in \texttt{try} blocks exercised by the test suite.\\
    \midrule
    
    \texttt{CATCH\_IC} & \textit{Catch Instruction Coverage}. The percentage of instructions in \texttt{catch} blocks exercised by the test suite.\\
    \midrule
    
    \texttt{CATCH\_BC} & \textit{Catch Branch Coverage}. The percentage of branches in \texttt{catch} blocks exercised by the test suite.\\
    \midrule
    
    \texttt{FINALLY\_IC} & \textit{Finally Instruction Coverage}. The percentage of instructions in \texttt{finally} blocks exercised by the test suite.\\
    \midrule
    
    \texttt{FINALLY\_BC} & \textit{Finally Branch Coverage}. The percentage of branches in \texttt{finally} blocks exercised by the test suite.\\
    \midrule

    \texttt{THROW\_IC} & \textit{Throw Instruction Coverage}. The percentage of \texttt{throw} instructions exercised by the test suite.\\
    \midrule
    
    \texttt{THROWS\_MC} & \textit{Throws Method Coverage}. The percentage of methods with a \texttt{throws} clause in its signature exercised by the test suite.\\
    \midrule
    \midrule
    
    \texttt{NON\_EH\_IC} & \textit{Non-Exception Handling Instruction Coverage}. The percentage of instructions exercised by the test suite that are not \texttt{throw} and not in \texttt{try}, \texttt{catch}, and \texttt{finally} blocks.\\
    \midrule
    
    \texttt{NON\_EH\_BC} & \textit{Non-Exception Handling Branch Coverage}. The percentage of branches exercised by the test suite that are not in \texttt{try}, \texttt{catch}, and \texttt{finally} blocks.\\
    \midrule
    
    \texttt{NON\_TRY\_IC} & \textit{Non-Try Instruction Coverage}. The percentage of instructions exercised by the test suite that are not in \texttt{try} blocks.\\
    \midrule

    \texttt{NON\_TRY\_BC} & \textit{Non-Try Branch Coverage}. The percentage of branches exercised by the test suite that are not in \texttt{try} blocks.\\
    \midrule
    
    \texttt{NON\_CATCH\_IC} & \textit{Non-Catch Instruction Coverage}. The percentage of instructions exercised by the test suite that are not in \texttt{catch} blocks.\\
    \midrule

    \texttt{NON\_CATCH\_BC} & \textit{Non-Catch Branch Coverage}. The percentage of branches exercised by the test suite that are not in \texttt{catch} blocks.\\
    \midrule
    
    \texttt{NON\_FINALLY\_IC} & \textit{Non-Finally Instruction Coverage}. The percentage of instructions exercised by the test suite that are not in \texttt{finally} blocks.\\
    \midrule
    
    \texttt{NON\_FINALLY\_BC} & \textit{Non-Finally Branch Coverage}. The percentage of branches exercised by the test suite that are not in \texttt{finally} blocks.\\
    \midrule
    
    \texttt{NON\_THROW\_IC} & \textit{Non-Throw Instruction Coverage}. The percentage of instructions exercised by the test suite that are not \texttt{throw}.\\
    \midrule

    \texttt{NON\_THROWS\_MC} & \textit{Non-Throws Method Coverage}. The percentage of methods exercised by the test suite without a \texttt{throws} clause in its signature.\\
    \bottomrule
    \end{tabular}
    }
\end{table}
\section{Study Results}\label{sec:results}

\subsection{Preliminary Observation of the Libraries' Overall Coverage}

To properly assess EH testing adequacy in terms of EH code coverage, we need to observe the libraries' overall coverage to serve as a point of comparison.
Otherwise, any high (or low) levels of EH code coverage that we observe in a library may be due to the high (or low) levels of overall coverage in the library. Thus, this serves as baseline that we can take into account when drawing conclusions from our observations.

\begin{figure}
	\centering
	\includegraphics[scale=0.80]{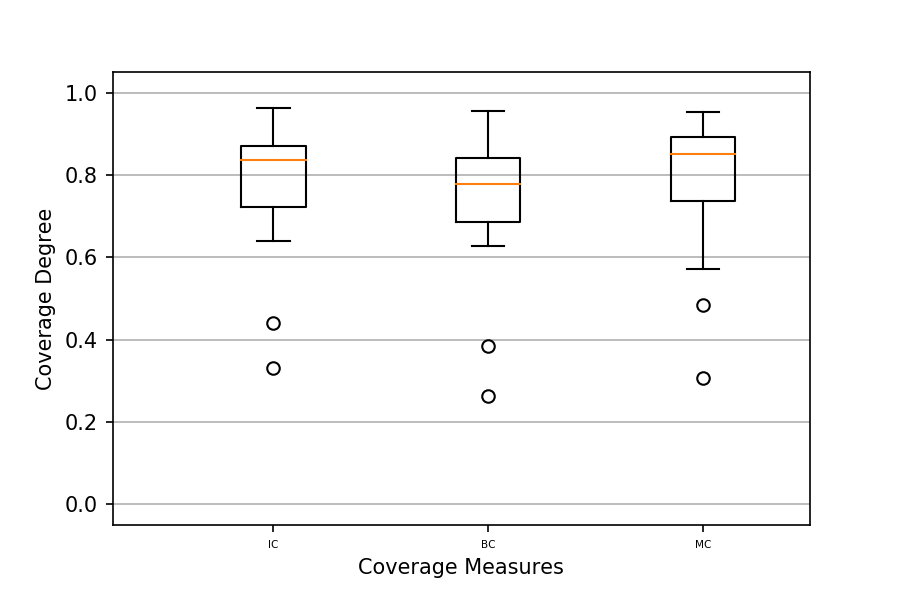}
	\caption{Overall code coverage boxplots of the 27 studied libraries (\texttt{IC} - Instruction Coverage, \texttt{BC} - Branch Coverage, and \texttt{MC} - Method Coverage).}
	\label{fig:AllCov}
\end{figure}

Fig.~\ref{fig:AllCov} presents boxplots depicting the distribution of overall instruction, branch, and method coverage, as detailed in Section~\ref{sec:CodeCoverage} and Table~\ref{tab:GeneralCodeCoverage}.
Note that the distributions were computed considering all the 27 studied libraries.
The median values for Instruction Coverage (\texttt{IC}), Branch Coverage (\texttt{BC}) and Method Coverage (\texttt{MC}) are 82\%, 78\% and 83\%, respectively.
One must notice that apart from 2 outliers, all studied libraries tend to present coverage degrees in medium to high echelons, reaching more than 95\% of coverage for some libraries in all metrics.
This indicates that the libraries under study exhibit mature testing practices for the libraries' overall source code.


\subsection{RQ1. What is the test coverage of EH code in long-lived Java libraries?}\label{sec:RQ1Results}

\begin{center}
\cornersize{.2}
\Ovalbox{\begin{minipage}{.95\textwidth}
\textbf{Summary of RQ1:} \texttt{try} and \texttt{finally} blocks are largely more covered than \texttt{catch} blocks and \texttt{throw} statements, indicating that the test suites are struggling to raise and test exceptional behaviors in the programs.
\end{minipage}}
\end{center}

Table~\ref{tab:ComputedGeneralCodeCoverage} presents the computed code coverage metrics of the first three sections of Table~\ref{tab:GeneralCodeCoverage} for all libraries included in this study.Not all metrics could be computed for all libraries. 
For instance, the BeanUtils library presents no \texttt{finally} block in its source code.
As a result, all \texttt{finally}-related coverage metrics (\texttt{FINALLY\_IC} and \texttt{FINALLY\_BC}) could not be computed.
We indicate with a `-' all cases in which a certain coverage metric could not be computed for a certain library.

\begin{table}[!htbp]
\footnotesize
\renewcommand{\arraystretch}{1.2}
\caption{Computed general code coverage measures per library.}
\centering
\label{tab:ComputedGeneralCodeCoverage}
\begin{adjustbox}{angle=90}
\begin{tabular}{l|r|r|r|r|r|r|r|r|r|r|r|r|r|r}
    \toprule
    \multicolumn{1}{c|}{\textbf{Library}} & 
    \multicolumn{1}{c|}{\texttt{IC}} & 
    \multicolumn{1}{c|}{\texttt{BC}} &
    \multicolumn{1}{c|}{\texttt{MC}} &
    \multicolumn{1}{c|}{\texttt{TRY\_CATCH\_BC}} &    
    \multicolumn{1}{c|}{\texttt{EH\_IC}} &
    \multicolumn{1}{c|}{\texttt{EH\_BC}} &
    \multicolumn{1}{c|}{\texttt{TRY\_IC}} & 
    \multicolumn{1}{c|}{\texttt{TRY\_BC}} & 
    \multicolumn{1}{c|}{\texttt{CATCH\_IC}} &
    \multicolumn{1}{c|}{\texttt{CATCH\_BC}} & 
    \multicolumn{1}{c|}{\texttt{FINALLY\_IC}} & 
    \multicolumn{1}{c|}{\texttt{FINALLY\_BC}} &
    \multicolumn{1}{c|}{\texttt{THROW\_IC}} &
    \multicolumn{1}{c}{\texttt{THROWS\_MC}} \\
    \midrule
    \textbf{BCEL}          & 0.44 & 0.38 & 0.48 & 0.02 & 0.31 & 0.33 & 0.38 & 0.35 & 0.13 & 0.25 & 0.15 & 0.11 & 0.01 & 0.70\\
    \textbf{BeanUtils}     & 0.65 & 0.65 & 0.68 & 0.30 & 0.52 & 0.48 & 0.68 & 0.65 & 0.32 & 0.26 & - & - & 0.43 & 0.78\\
    \textbf{CLI}           & 0.96 & 0.93 & 0.94 & 0.73 & 0.97 & 1.00 & 1.00 & 1.00 & 1.00 & - & 1.00 & - & 0.90 & 1.00\\
    \textbf{Codec}         & 0.96 & 0.91 & 0.90 & 0.55 & 0.78 & 0.97 & 0.84 & 0.97 & 0.50 & - & 0.88 & - & 0.62 & 0.87\\
    \textbf{Collections}   & 0.87 & 0.81 & 0.87 & 0.33 & 0.68 & 0.68 & 0.88 & 0.63 & 0.63 & 1.00 & 0.57 & 1.00 & 0.64 & 0.94\\
    \textbf{Compress}      & 0.85 & 0.76 & 0.85 & 0.22 & 0.66 & 0.75 & 0.86 & 0.78 & 0.18 & 0.25 & 0.91 & 0.68 & 0.30 & 0.93\\
    \textbf{Configuration} & 0.88 & 0.84 & 0.91 & 0.50 & 0.76 & 0.74 & 0.81 & 0.77 & 0.45 & 0.42 & 1.00 & 0.70 & 0.72 & 0.93\\
    \textbf{DBCP}          & 0.70 & 0.70 & 0.92 & 0.04 & 0.55 & 0.68 & 0.92 & 0.71 & 0.02 & 0.13 & 0.86 & 0.75 & 0.20 & 0.94\\
    \textbf{DbUtils}       & 0.64 & 0.77 & 0.57 & 0.29 & 0.66 & 0.77 & 0.82 & 0.71 & 0.07 & - & 0.93 & 0.88 & 0.41 & 0.34\\
    \textbf{Digester}      & 0.66 & 0.65 & 0.74 & 0.16 & 0.57 & 0.53 & 0.83 & 0.67 & 0.18 & 0.13 & 1.00 & 0.50 & 0.22 & 0.86\\
    \textbf{Email}         & 0.72 & 0.67 & 0.81 & 0.16 & 0.65 & 0.57 & 0.77 & 0.67 & 0.59 & 0.75 & 0.29 & 0.15 & 0.24 & 0.87\\
    \textbf{Exec}          & 0.72 & 0.63 & 0.74 & 0.10 & 0.42 & 0.54 & 0.43 & 0.52 & 0.27 & 0.50 & 0.75 & 0.67 & 0.31 & 0.65\\
    \textbf{FileUpload}    & 0.80 & 0.76 & 0.67 & 0.26 & 0.69 & 0.69 & 0.79 & 0.79 & 0.24 & - & 0.81 & 0.50 & 0.44 & 0.68\\
    \textbf{Functor}       & 0.82 & 0.66 & 0.90 & - & 0.23 & - & - & - & - & - & - & - & 0.23 & - \\
    \textbf{IO}            & 0.90 & 0.88 & 0.89 & 0.42 & 0.78 & 0.78 & 0.91 & 0.79 & 0.18 & 0.75 & 0.82 & 0.70 & 0.74 & 0.91\\
    \textbf{Lang}          & 0.96 & 0.91 & 0.95 & 0.65 & 0.85 & 0.85 & 0.93 & 0.86 & 0.82 & 0.70 & 1.00 & 1.00 & 0.72 & 0.97\\
    \textbf{Math}          & 0.92 & 0.85 & 0.87 & 0.47 & 0.65 & 0.72 & 0.92 & 0.91 & 0.53 & 0.53 & 0.73 & - & 0.59 & 0.89\\
    \textbf{Net}           & 0.33 & 0.26 & 0.31 & 0.07 & 0.13 & 0.14 & 0.16 & 0.16 & 0.02 & 0.03 & 0.10 & 0.00 & 0.10 & 0.14\\
    \textbf{Pool}          & 0.84 & 0.79 & 0.89 & 0.51 & 0.85 & 0.86 & 0.93 & 0.87 & 0.55 & 0.75 & 0.98 & 1.00 & 0.71 & 0.95\\
    \textbf{Proxy}         & 0.82 & 0.80 & 0.85 & 0.48 & 0.55 & 0.43 & 0.52 & 0.43 & 1.00 & - & - & - & 0.62 & 1.00\\
    \textbf{Validator}     & 0.86 & 0.76 & 0.81 & 0.30 & 0.52 & 0.54 & 0.75 & 0.60 & 0.17 & 0.30 & 0.00 & - & 0.43 & 0.86\\
    \midrule
    \textbf{Gson}          & 0.84 & 0.79 & 0.85 & 0.28 & 0.67 & 0.66 & 0.83 & 0.63 & 0.47 & 1.00 & 1.00 & 1.00 & 0.34 & 0.97\\
    \textbf{Hamcrest}      & 0.83 & 0.95 & 0.71 & 0.22 & 0.66 & 1.00 & 1.00 & 1.00 & 0.55 & - & - & - & 0.26 & 1.00\\
    \textbf{Jsoup}         & 0.84 & 0.78 & 0.85 & 0.38 & 0.73 & 0.75 & 0.87 & 0.86 & 0.70 & - & 0.00 & 0.00 & 0.30 & 0.88\\
    \textbf{JUnit}         & 0.86 & 0.83 & 0.88 & 0.44 & 0.66 & 0.64 & 0.83 & 0.68 & 0.45 & 0.45 & 0.93 & 0.88 & 0.54 & 0.94\\
    \textbf{Mockito}       & 0.87 & 0.86 & 0.89 & 0.49 & 0.83 & 0.83 & 0.93 & 0.85 & 0.68 & 0.67 & 1.00 & 1.00 & 0.62 & 0.90\\
    \textbf{X-Stream}      & 0.78 & 0.74 & 0.77 & 0.17 & 0.65 & 0.72 & 0.83 & 0.75 & 0.13 & 0.17 & 0.87 & 0.50 & 0.14 & 0.65\\
    \bottomrule
    \end{tabular}
   \end{adjustbox}
\end{table}

\begin{figure}
	\centering
	\includegraphics[scale=0.80]{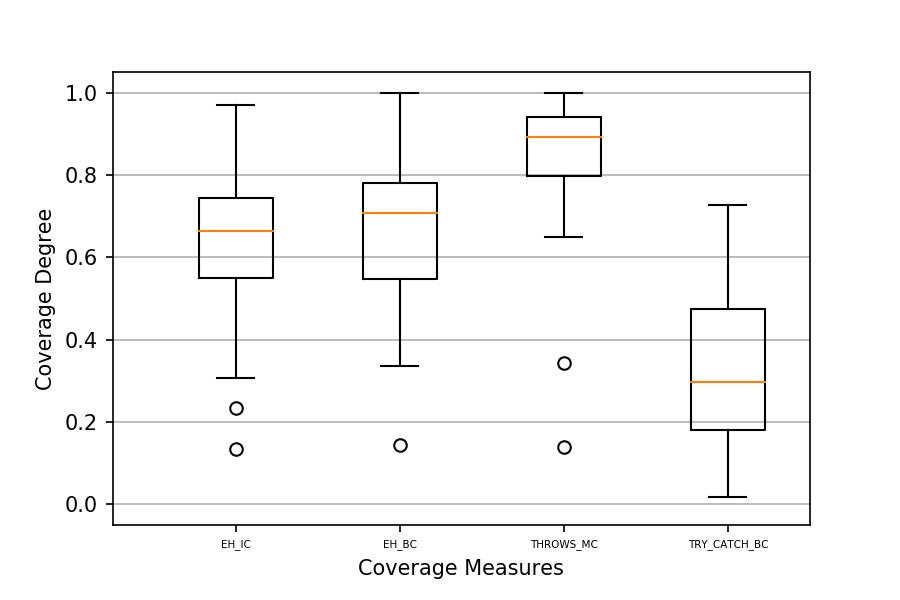}
	\caption{Distribution of general EH-related code coverage metrics for the libraries under study.}
	\label{fig:EHAllCov}
\end{figure}

The boxplots in Fig.~\ref{fig:EHAllCov} show the distribution of general coverage metrics for EH-related code. 
These are the EH coverage metrics that correspond to the overall coverage metrics displayed in Fig.~\ref{fig:AllCov} plus the \texttt{TRY\_CATCH\_BC} metric.
The coverage degree of \texttt{EH\_IC} ranges from 55\% to 74\%, \texttt{EH\_BC} ranges from 54\% to 78\%, \texttt{THROWS\_MC} ranges from 79\% to 94\%, and \texttt{TRY\_CATCH\_BC} ranges from 18\% to 47\%. 
Additionally, one should notice there exist libraries with 100\% coverage for \texttt{EH\_BC} and \texttt{THROWS\_MC} and with about 2\% for \texttt{TRY\_CATCH\_BC}.

We can draw interesting observations when comparing the EH-related coverage with the equivalent coverage metrics for the whole library displayed in Fig.~\ref{fig:AllCov}.
First, we observe a larger deviation in the adequacy of EH testing than in overall testing.
This is depicted by how the boxplots for EH coverage tend to be less compact than the overall ones, which tend to indicate that the EH testing practices tend to be less mature than the overall testing ones.
When considering instruction coverage for EH code, for example, we see libraries with less than 40\% of their EH instructions being covered, where the smallest overall instruction coverage is above 60\%.
Nevertheless, this is not always the case. 
We observed that a few libraries reached 100\% EH method coverage, which did not occur for overall method coverage in any library.
Particularly, looking at the \texttt{TRY\_CATCH\_BC} boxplot, we can also observe that coverage of \texttt{catch} blocks (i.e., the reachability of these blocks instead of the instructions or branches within them) is very low, indicating that the test suites of the studied libraries are not able to exercise the code derived from exceptional control flows.

\begin{figure}[!h]
	\centering
	\includegraphics[scale=0.80]{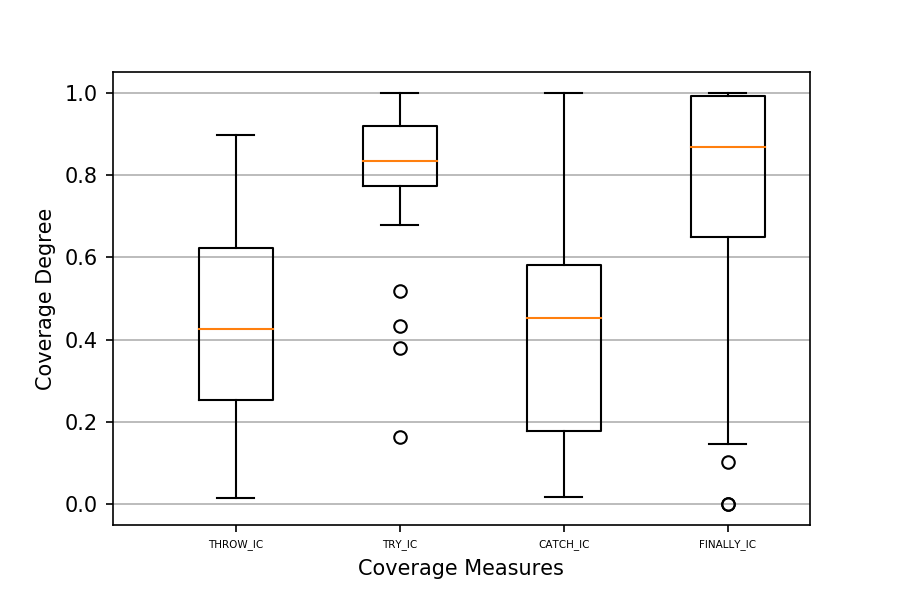}
	\caption{EH instruction code coverage boxplots of studied libraries.}
	\label{fig:ICov}
\end{figure}

\begin{figure}[!h]
	\centering
	\includegraphics[scale=0.80]{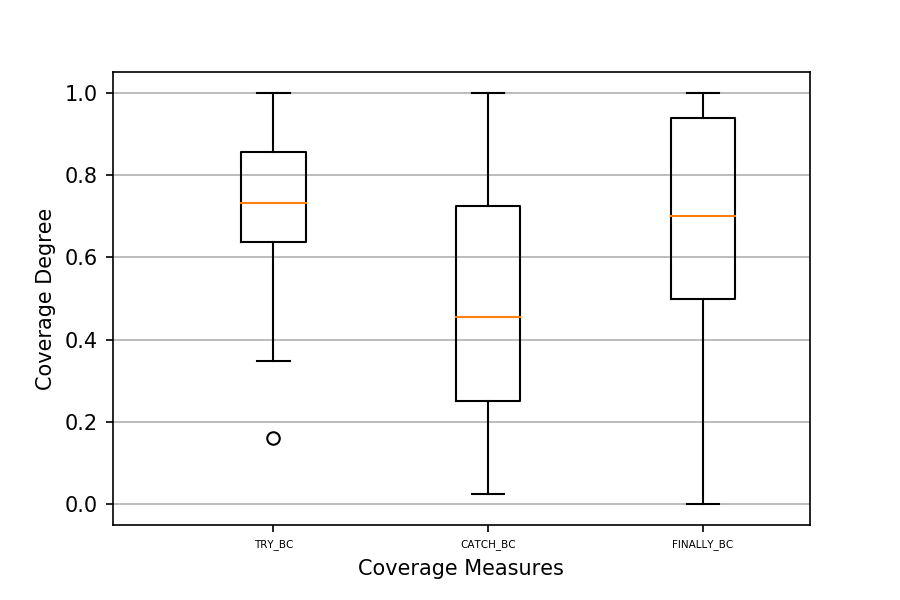}
	\caption{EH branch code coverage boxplots of studied libraries.}
	\label{fig:BCov}
\end{figure}

We also plotted boxplots detailing the internal distribution of \texttt{EH\_IC} (see Fig.~\ref{fig:ICov}) and \texttt{EH\_BC} (see Fig.~\ref{fig:BCov}). Looking at Fig.~\ref{fig:ICov} and the data in Table~\ref{tab:ComputedGeneralCodeCoverage}, one can see that instructions in \texttt{try} and \texttt{finally} blocks have the best coverage degrees. 
In fact, they assume high levels of coverage if one consider the interquartile interval, ranging from 77\% to 91\% for \texttt{TRY\_IC} and from 64\% to 99\% for \texttt{FINALLY\_IC}. 
Differently, when considering the lowest quartile, the \texttt{throw} instructions and \texttt{catch} blocks have the worst coverage, with 25\% and 17\%, respectively. 
Hence, this suggests that \texttt{THROW\_IC} and \texttt{CATCH\_IC} are the metrics that impact the general \texttt{EH\_IC} the most.

Looking at Fig.~\ref{fig:BCov} and the data in Table~\ref{tab:ComputedGeneralCodeCoverage}, one can see that the branches in \texttt{try} and \texttt{finally} blocks have the better coverage when compared to the branches in \texttt{catch} blocks. 
In fact, if one consider the median of \texttt{TRY\_BC} (73\%) and \texttt{FINALLY\_BC} (70\%), one will see that about three-quarters of \texttt{CATCH\_BC} is covered less than the median coverage of \texttt{TRY\_BC} and \texttt{FINALLY\_BC}.
Thus, this suggests that \texttt{CATCH\_BC} coverage is the one that impact most of the \texttt{EH\_BC} coverage.

When analyzing the details of both instruction and branch coverage for EH code, we find a similar pattern, where \texttt{try} and \texttt{finally} blocks are largely more covered than \texttt{throw} instructions and \texttt{catch} blocks.
This is a worrisome observation because \texttt{try} and \texttt{finally} blocks are always executed in non-exceptional behaviors. Hence, we deduct that the test suites of the studied libraries are failing to raise internal (coded in the library) and external (signaled from third-party libraries as a return of a method call) exceptions. Thus, despite presenting high coverage for instruction and branches in the overall source code and EH code, the tests are still mostly exercising non-exceptional flows within the programs, where the exceptional control flows are not being well tested. This is supported by the results of the \texttt{TRY\_CATCH\_BC} metric.

\subsection{RQ2. What is the difference between EH and non-EH code coverage in long-lived Java libraries?}\label{sec:RQ2Results}

\begin{center}
\cornersize{.2}
\Ovalbox{\begin{minipage}{.95\textwidth}
\textbf{Summary of RQ2:} EH code is significantly less covered by test suites than non-EH code, especially regarding instructions and branches within \texttt{catch} blocks and \texttt{throw} instructions.
\end{minipage}}
\end{center}

To answer this research question, we first computed the complementary code coverage metrics (see Table~\ref{tab:GeneralCodeCoverage}) for each studied library and summarize them in Table~\ref{tab:ComputedComplementaryCodeCoverage}. 
Next, we employ two statistical tests to compare EH and non-EH code coverage values, in which we can verify whether there are statistically significant differences between them. The first test is the Kolmogorov–Smirnov test (KS), and the second is the Mann-Whitney test (MW), as detailed in the next paragraphs. We also compute the effect size using Cliff's delta statistic~\citep{Cliff1993}, a non-parametric measure that quantifies the amount of difference between two groups of observations (EH and non-EH code coverage). This measure can be seen as a complementary analysis for the corresponding hypothesis testing and p-values. We display the values for the Cliff's delta measure and their respective interpretation in the last column of Table~\ref{tab:HypothesisTest}. 

\begin{table}
\footnotesize
\renewcommand{\arraystretch}{1.2}
\caption{Computed complementary code coverage measures per library.}
\centering
\label{tab:ComputedComplementaryCodeCoverage}
\begin{adjustbox}{angle=90}
\begin{tabular}{l|r|r|r|r|r|r|r|r|r|r}
    \toprule
    \multicolumn{1}{c|}{\textbf{Library}} & 
    \multicolumn{1}{c|}{\texttt{NON\_EH\_IC}} & 
    \multicolumn{1}{c|}{\texttt{NON\_EH\_BC}} &
    \multicolumn{1}{c|}{\texttt{NON\_TRY\_IC}} & 
    
    \multicolumn{1}{c|}{\texttt{NON\_TRY\_BC}} & 
    \multicolumn{1}{c|}{\texttt{NON\_CATCH\_IC}} &
    \multicolumn{1}{c|}{\texttt{NON\_CATCH\_BC}} & 
    
    \multicolumn{1}{c|}{\texttt{NON\_FINALLY\_IC}} & 
    \multicolumn{1}{c|}{\texttt{NON\_FINALLY\_BC}} &
    \multicolumn{1}{c|}{\texttt{NON\_THROW\_IC}} &
    
    \multicolumn{1}{c}{\texttt{NON\_THROWS\_MC}} \\
    \midrule
    \textbf{BCEL} & 0.45 & 0.04 & 0.44 & 0.39 & 0.44 & 0.38 & 0.44 & 0.39 & 0.45 & 0.47\\
    \textbf{BeanUtils} & 0.66 & 0.16 & 0.64 & 0.65 & 0.66 & 0.67 & 0.65 & 0.65 & 0.65 & 0.66\\
    \textbf{CLI} & 0.96 & 0.76 & 0.96 & 0.93 & 0.96 & 0.93 & 0.96 & 0.93 & 0.96 & 0.93\\
    \textbf{Codec} & 0.97 & 0.54 & 0.96 & 0.91 & 0.96 & 0.91 & 0.96 & 0.91 & 0.97 & 0.91\\
    \textbf{Collections} & 0.87 & 0.41 & 0.87 & 0.81 & 0.87 & 0.81 & 0.87 & 0.81 & 0.87 & 0.87\\
    \textbf{Compress} & 0.85 & 0.28 & 0.85 & 0.76 & 0.85 & 0.76 & 0.85 & 0.76 & 0.85 & 0.82\\
    \textbf{Configuration} & 0.89 & 0.38 & 0.88 & 0.84 & 0.88 & 0.84 & 0.88 & 0.84 & 0.88 & 0.91\\
    \textbf{DBCP} & 0.78 & 0.13 & 0.64 & 0.70 & 0.80 & 0.71 & 0.69 & 0.70 & 0.71 & 0.89\\
    \textbf{DbUtils} & 0.63. & 0.10 & 0.62 & 0.78 & 0.65 & 0.77 & 0.63 & 0.77 & 0.64 & 0.84\\
    \textbf{Digester} & 0.67 & 0.14 & 0.65 & 0.65 & 0.67 & 0.65 & 0.66 & 0.65 & 0.67 & 0.72\\
    \textbf{Email} & 0.74 & 0.19 & 0.72 & 0.67 & 0.72 & 0.67 & 0.73 & 0.69 & 0.74 & 0.77\\
    \textbf{Exec} & 0.78 & 0.21 & 0.75 & 0.64 & 0.73 & 0.63 & 0.72 & 0.63 & 0.73 & 0.75\\
    \textbf{FileUpload} & 0.81 & 0.25 & 0.80 & 0.76 & 0.80 & 0.76 & 0.80 & 0.77 & 0.80 & 0.67\\
    \textbf{Functor} & 0.83 & 0.38 & 0.82 & 0.66 & 0.82 & 0.66 & 0.82 & 0.66 & 0.83 & 0.90\\
    \textbf{IO} & 0.92 & 0.50 & 0.90 & 0.89 & 0.91 & 0.88 & 0.90 & 0.88 & 0.91 & 0.88\\
    \textbf{Lang} & 0.96 & 0.75 & 0.96 & 0.91 & 0.96 & 0.91 & 0.96 & 0.91 & 0.96 & 0.95\\
    \textbf{Math} & 0.93 & 0.41 & 0.92 & 0.85 & 0.93 & 0.85 & 0.92 & 0.85 & 0.93 & 0.87\\
    \textbf{Net} & 0.36 & 0.03 & 0.35 & 0.28 & 0.34 & 0.27 & 0.33 & 0.26 & 0.33 & 0.40\\
    \textbf{Pool} & 0.84 & 0.26 & 0.83 & 0.78 & 0.85 & 0.79 & 0.84 & 0.79 & 0.85 & 0.88\\
    \textbf{Proxy} & 0.84 & 0.22 & 0.84 & 0.82 & 0.82 & 0.80 & 0.82 & 0.80 & 0.83 & 0.84\\
    \textbf{Validator} & 0.87 & 0.31 & 0.86 & 0.76 & 0.87 & 0.76 & 0.86 & 0.76 & 0.87 & 0.81\\
    \textbf{Gson} & 0.85 & 0.36 & 0.84 & 0.80 & 0.84 & 0.79 & 0.83 & 0.79 & 0.85 & 0.83\\
    \midrule
    \textbf{Hamcrest} & 0.84 & 0.24 & 0.83 & 0.95 & 0.84 & 0.95 & 0.83 & 0.95 & 0.84 & 0.71\\
    \textbf{Jsoup} & 0.84 & 0.35 & 0.84 & 0.78 & 0.84 & 0.78 & 0.84 & 0.78 & 0.84 & 0.85\\
    \textbf{JUnit} & 0.87 & 0.30 & 0.86 & 0.83 & 0.87 & 0.83 & 0.86 & 0.83 & 0.86 & 0.88\\
    \textbf{Mockito} & 0.88 & 0.34 & 0.87 & 0.86 & 0.87 & 0.86 & 0.87 & 0.86 & 0.88 & 0.89\\
    \textbf{X-Stream} & 0.79 & 0.25 & 0.77 & 0.74 & 0.78 & 0.75 & 0.78 & 0.74 & 0.79 & 0.78\\
    \bottomrule
    \end{tabular}
   \end{adjustbox}
\end{table}

The KS is a two-sided test for the null hypothesis that two independent samples are drawn from the same continuous distribution. 
We test this null hypothesis by taking into account pairs of samples of non-EH and EH coverage metrics (see Tables \ref{tab:GeneralCodeCoverage}, \ref{tab:ComputedGeneralCodeCoverage} and \ref{tab:ComputedComplementaryCodeCoverage}). 
Consider instruction coverage, for example, where we abbreviate it to simply $A$ for brevity.
We measured both \texttt{EH\_IC} and \texttt{NON\_EH\_IC} for all 27 libraries under study.
We formulate our null hypothesis as $\mathcal{H}_{0}^{A}$: \texttt{NON\_EH\_IC} $=$ \texttt{EH\_IC}. 
In case this KS null hypothesis cannot be rejected, we assume that non-EH and EH code coverage measures have the same distribution, i.e., there is no statistical difference in instruction coverage for EH and non-EH code when considereing all libraries.
However, in case the KS null hypothesis is rejected, we assume the alternative hypothesis $\mathcal{H}_{1}^{A}$:~\texttt{NON\_EH\_IC} $\neq$ \texttt{EH\_IC}, which indicates statistical difference in instruction coverage between EH and non-EH code.

In case statistical difference is indicated by the KS test, we can employ the MW test to assert whether the coverage values in non-EH code are higher than the coverage values in EH code, or vice-versa. It is important to observe that the MW test is only performed if the null hypothesis of KS is rejected. 
Consider the instruction coverage metric, for example.
The MW test considers the null hypothesis $\mathcal{H}_{0}^{A}$:~\texttt{NON\_EH\_IC} $>$ \texttt{EH\_IC}. 
If the MW null hypothesis cannot be rejected, we assume that the instruction coverage of non-EH code is significantly greater than the instruction coverage in EH code. 
Otherwise, we assume MW's alternative hypothesis ($\mathcal{H}_{1}^{A}$: \texttt{NON\_EH\_IC} $<$ \texttt{EH\_IC}), which indicates that instruction coverage in EH code is significantly greater than in non-EH code.
Table~\ref{tab:HypothesisTest} presents the statistical tests results for all coverage metrics with the significance level of $\alpha < 0.05$.

\begin{table}
\renewcommand{\arraystretch}{1.30}
\caption{Summary of hypothesis statement, the statistics test, and the Cliff's Delta effect size results. The symbols {\color{blue}\checkmark} and {\color{red}\xmark} indicate the result of the null hypothesis test ({\color{blue}\checkmark} fail to reject, and {\color{red}\xmark} reject). The Cliff's Delta effect size interpretation: negligible $= [0, 0.147)$, small $= [0.147, 0.33)$, medium $= [0.33, 0.474)$, and large $= [0.474, 1]$. \textcolor{blue}{The symbol $\ast$ indicates the p-value adjustment applying the~\cite{Benjamini2001}'s method on the pairs of p-values obtained by the two consecutive statistical tests, KS and MW, per line. The symbol $\dagger$ indicates the p-value adjustment applying the~\cite{Benjamini2001}'s method on the sets of p-values where the metrics involved in each the test (per column) are not independent.}}
\centering
\label{tab:HypothesisTest}
\footnotesize
\resizebox{\columnwidth}{!}{
\begin{tabular}{l|r|l|c|c}
    \toprule
    \multicolumn{1}{c}{\textbf{KS Hypothesis}} & \multicolumn{1}{|c}{\textbf{p-value}} & \multicolumn{1}{|c}{\textbf{MW Hypothesis}} & \multicolumn{1}{|c}{\textbf{p-value}} & \multicolumn{1}{|c}{\textbf{|Effect Size|}}\\
    \midrule
    $\mathcal{H}_{0}^{A}$: \texttt{NON\_EH\_IC} $=$    \texttt{EH\_IC} ({\color{red}\xmark}) & \multirow{2}{*}{$5.7 \times 10^{-4}{}^{\ast\dagger}$} & $\mathcal{H}_{0}^{A}$: \texttt{NON\_EH\_IC} $>$ \texttt{EH\_IC} ({\color{blue}\checkmark}) & \multirow{2}{*}{$2.8 \times 10^{-4}{}^{\ast\dagger}$} & $0.61$\\
    $\mathcal{H}_{1}^{A}$: \texttt{NON\_EH\_IC} $\neq$ \texttt{EH\_IC} & & $\mathcal{H}_{1}^{A}$: \texttt{NON\_EH\_IC} $<$ \texttt{EH\_IC} & & (large)\\
    \midrule

    $\mathcal{H}_{0}^{B}$: \texttt{NON\_EH\_BC} $=$    \texttt{EH\_BC} ({\color{red}\xmark}) & \multirow{2}{*}{$5.3 \times 10^{-7}{}^\ast\dagger$} & $\mathcal{H}_{0}^{B}$: \texttt{NON\_EH\_BC} $>$ \texttt{EH\_BC} ({\color{red}\xmark}) & \multirow{2}{*}{$1.4 \times 10^{-6}{}^{\ast\dagger}$} & $0.80$\\
    $\mathcal{H}_{1}^{B}$: \texttt{NON\_EH\_BC} $\neq$ \texttt{EH\_BC} & & $\mathcal{H}_{1}^{B}$: \texttt{NON\_EH\_BC} $<$ \texttt{EH\_BC} & & (large)\\
    \midrule

    $\mathcal{H}_{0}^{C}$: \texttt{NON\_THROWS\_MC} $=$    \texttt{THROWS\_MC} ({\color{red}\xmark}) & \multirow{2}{*}{$4.5 \times 10^{-2}{}^\ast$} & $\mathcal{H}_{0}^{C}$: \texttt{NON\_THROWS\_MC} $>$ \texttt{THROWS\_MC} ({\color{red}\xmark}) & \multirow{2}{*}{$4.5 \times 10^{-2}{}^\ast$} & $0.30$\\
    $\mathcal{H}_{1}^{C}$: \texttt{NON\_THROWS\_MC} $\neq$ \texttt{THROWS\_MC} & & $\mathcal{H}_{1}^{C}$: \texttt{NON\_THROWS\_MC} $<$ \texttt{THROWS\_MC} & & (small)\\
    \midrule
    
    $\mathcal{H}_{0}^{D}$: \texttt{NON\_THROW\_IC} $=$    \texttt{THROW\_IC} ({\color{red}\xmark}) & \multirow{2}{*}{$3.1 \times 10^{-6}{}^{\ast\dagger}$} & $\mathcal{H}_{0}^{D}$: \texttt{NON\_THROW\_IC} $>$ \texttt{THROW\_IC} ({\color{blue}\checkmark}) & \multirow{2}{*}{$1.5 \times 10^{-6}{}^{\ast\dagger}$} & $0.82$\\
    $\mathcal{H}_{1}^{D}$: \texttt{NON\_THROW\_IC} $\neq$ \texttt{THROW\_IC} &  & $\mathcal{H}_{1}^{D}$: \texttt{NON\_THROW\_IC} $<$ \texttt{THROW\_IC} & & (large)\\
    \midrule
    
    $\mathcal{H}_{0}^{E}$: \texttt{NON\_TRY\_IC} $=$    \texttt{TRY\_IC} ({\color{blue}\checkmark}) & \multirow{2}{*}{$1.0^{\dagger}$}     & $\mathcal{H}_{0}^{E}$: \texttt{NON\_TRY\_IC} $>$ \texttt{TRY\_IC} & \multirow{2}{*}{N/A} & $0.06$\\
    $\mathcal{H}_{1}^{E}$: \texttt{NON\_TRY\_IC} $\neq$ \texttt{TRY\_IC} & & $\mathcal{H}_{1}^{E}$: \texttt{NON\_TRY\_IC} $<$ \texttt{TRY\_IC} & & (negligible)\\
    \midrule
    
    $\mathcal{H}_{0}^{F}$: \texttt{NON\_CATCH\_IC} $=$    \texttt{CATCH\_IC} ({\color{red}\xmark}) & \multirow{2}{*}{$5.0 \times 10^{-6}{}^{\ast\dagger}$} & $\mathcal{H}_{0}^{F}$: \texttt{NON\_CATCH\_IC} $>$ \texttt{CATCH\_IC} ({\color{blue}\checkmark}) & \multirow{2}{*}{$8.8 \times 10^{-6}{}^{\ast\dagger}$} & $0.74$\\
    $\mathcal{H}_{1}^{F}$: \texttt{NON\_CATCH\_IC} $\neq$ \texttt{CATCH\_IC} & & $\mathcal{H}_{1}^{F}$: \texttt{NON\_CATCH\_IC} $<$ \texttt{CATCH\_IC} & & (large)\\
    \midrule
    
    $\mathcal{H}_{0}^{G}$: \texttt{NON\_FINALLY\_IC} $=$    \texttt{FINALLY\_IC} ({\color{blue}\checkmark}) & \multirow{2}{*}{$4.6 \times 10^{-1}{}^{\dagger}$} & $\mathcal{H}_{0}^{G}$: \texttt{NON\_FINALLY\_IC} $>$ \texttt{FINALLY\_IC} & \multirow{2}{*}{N/A} & $0.16$\\
    $\mathcal{H}_{1}^{G}$: \texttt{NON\_FINALLY\_IC} $\neq$ \texttt{FINALLY\_IC} & & $\mathcal{H}_{1}^{G}$: \texttt{NON\_FINALLY\_IC} $<$ \texttt{FINALLY\_IC} & & (small)\\
    \midrule    
    
    $\mathcal{H}_{0}^{H}$: \texttt{NON\_TRY\_BC} $=$    \texttt{TRY\_BC} ({\color{blue}\checkmark}) & \multirow{2}{*}{$1.0^{\dagger}$}     & $\mathcal{H}_{0}^{H}$: \texttt{NON\_TRY\_BC} $>$ \texttt{TRY\_BC}    & \multirow{2}{*}{N/A} & $0.12$\\
    $\mathcal{H}_{1}^{H}$: \texttt{NON\_TRY\_BC} $\neq$ \texttt{TRY\_BC} & & $\mathcal{H}_{1}^{H}$: \texttt{NON\_TRY\_BC} $<$ \texttt{TRY\_BC} & & (negligible)\\
    \midrule
    
    $\mathcal{H}_{0}^{I}$: \texttt{NON\_CATCH\_BC} $=$    \texttt{CATCH\_BC} ({\color{red}\xmark}) & \multirow{2}{*}{$5.4 \times 10^{-3}{}^{\ast\dagger}$} & $\mathcal{H}_{0}^{I}$: \texttt{NON\_CATCH\_BC} $>$ \texttt{CATCH\_BC} ({\color{blue}\checkmark}) & \multirow{2}{*}{$1.4 \times 10^{-3}{}^{\ast\dagger}$} & $0.59$\\
    $\mathcal{H}_{1}^{I}$: \texttt{NON\_CATCH\_BC} $\neq$ \texttt{CATCH\_BC} & & $\mathcal{H}_{1}^{I}$: \texttt{NON\_CATCH\_BC} $<$ \texttt{CATCH\_BC} & & (large)\\
    \midrule
    
    $\mathcal{H}_{0}^{J}$: \texttt{NON\_FINALLY\_BC} $=$    \texttt{FINALLY\_BC} ({\color{blue}\checkmark}) & \multirow{2}{*}{$4.4\times 10^{-1}{}^{\dagger}$} & $\mathcal{H}_{0}^{J}$: \texttt{NON\_FINALLY\_BC} $>$ \texttt{FINALLY\_BC} & \multirow{2}{*}{N/A} & $0.14$\\
    $\mathcal{H}_{1}^{J}$: \texttt{NON\_FINALLY\_BC} $\neq$ \texttt{FINALLY\_BC} & & $\mathcal{H}_{1}^{J}$: \texttt{NON\_FINALLY\_BC} $<$ \texttt{FINALLY\_BC} & & (negligible)\\
    \bottomrule
    
    \end{tabular}
    }
\end{table}

\begin{figure}
	\centering
	\includegraphics[scale=0.80]{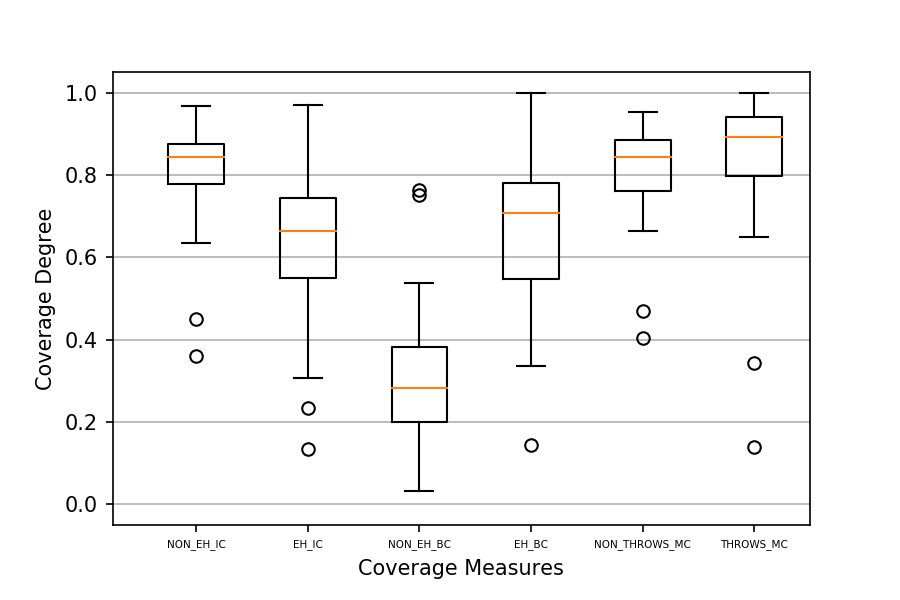}
	\caption{Overall EH and non-EH code coverage boxplots of studied libraries.}
	\label{fig:AllCovTest}
\end{figure}

To enhance this analysis, in Fig.~\ref{fig:AllCovTest}, we depict boxplots for the instruction, branch and method coverage for both non-EH and EH code.
When comparing the boxplots, one can see that the EH instruction coverage (\texttt{EH\_IC}) is lower than non-EH instruction coverage (\texttt{NON\_EH\_IC}). 
This perception is confirmed by the statistical tests results that reject the KS null hypothesis $\mathcal{H}_{0}^{A}$:~\texttt{NON\_EH\_IC} $=$ \texttt{EH\_IC} and did not reject the MW null hypothesis $\mathcal{H}_{0}^{A}$:~\texttt{NON\_EH\_IC} $>$ \texttt{EH\_IC}. 
This indicates that not only the instruction coverage of EH and non-EH code are statistically different but also that non-EH code is statistically more covered than EH code.

On the other hand, when considering branch coverage, EH code (\texttt{EH\_BC}) seems to be more covered than non-EH code (\texttt{NON\_EH\_BC}). 
Indeed, this perception is confirmed by the statistical tests results that reject the KS null hypothesis $\mathcal{H}_{0}^{B}$:~\texttt{NON\_EH\_BC} $=$ \texttt{EH\_BC} and also reject the MW null hypothesis $\mathcal{H}_{0}^{B}$:~\texttt{NON\_EH\_BC} $>$ \texttt{EH\_BC}. 
This is a counterintuitive observation given the results previously observed in our study.
We address this during our study' discussion (see Section~\ref{sec:discussion}).

Finally, different from instruction and branch coverage, the values of method coverage for EH code (\texttt{THROWS\_MC}) and non-EH code (\texttt{NON\_THROWS\_MC}) are visually similar.
However, this perception is not confirmed by the statistical tests that reject both the KS null hypothesis $\mathcal{H}_{0}^{C}$: \texttt{NON\_THROWS\_MC} $=$ \texttt{THROWS\_MC} and the MW null hypothesis $\mathcal{H}_{0}^{C}$: \texttt{NON\_THROWS\_MC} $>$ \texttt{THROWS\_MC}.
This indicates that methods without a \texttt{throws} clause are significantly less covered than methods with a \texttt{throws} clause.
Implications for this finding are also addressed in our discussion section.

\begin{figure}
	\centering
	\includegraphics[scale=0.80]{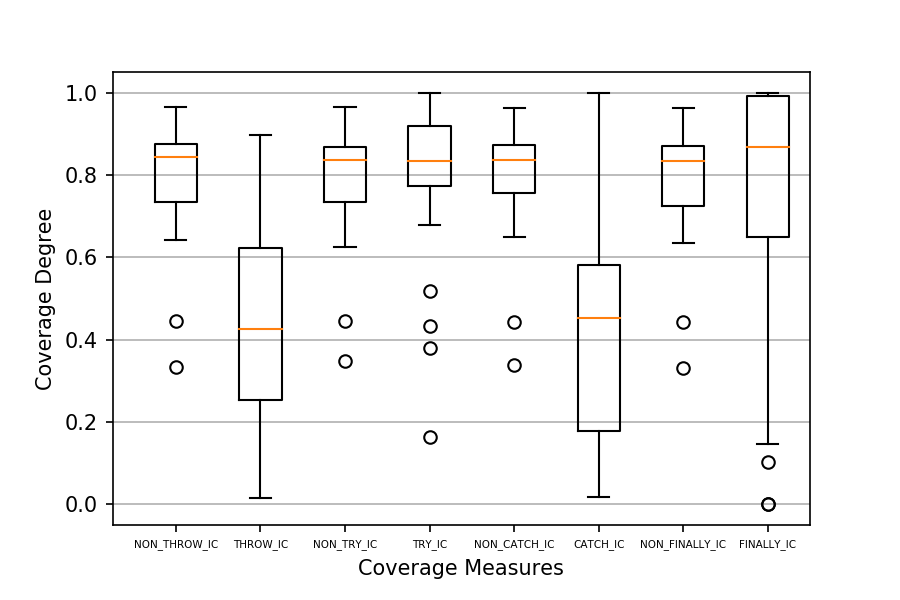}
     \caption{The EH and non-EH instruction coverage boxplots of studied libraries.}
	\label{fig:EHICTest}
\end{figure}

Fig.~\ref{fig:EHICTest} presents boxplots detailing instruction coverage metrics for non-EH and EH code.
It depicts coverage values for \texttt{throw} instructions and instructions in \texttt{try}, \texttt{catch} and \texttt{finally} blocks, respectively.
When comparing the boxplots of \texttt{THROW\_IC} and \texttt{NON\_THROW\_IC} metrics, one may notice that the \texttt{throw} instructions are less covered than the non-\texttt{throw} instructions. 
This perception is confirmed by the statistical tests results that reject the KS null hypothesis $\mathcal{H}_{0}^{D}$: \texttt{NON\_THROW\_IC} $=$ \texttt{THROW\_IC} and accept the MW null hypothesis $\mathcal{H}_{0}^{D}$:~\texttt{NON\_THROW\_IC} $>$ \texttt{THROW\_IC}. 
This indicates that even though the libraries under study present a fairly high level of instruction coverage (see Fig.~\ref{fig:AllCov}), the instructions that actually raise exceptions are not well covered.

When comparing the boxplots of \texttt{CATCH\_IC} and \texttt{NON\_CATCH\_IC} coverage, one can see that the instructions inside \texttt{catch} blocks are covered less than the instructions outside \texttt{catch} blocks. 
Once again, the statistical tests results confirm this perception by rejecting the KS null hypothesis $\mathcal{H}_{0}^{F}$:~\texttt{NON\_CATCH\_IC} $=$ \texttt{CATCH\_IC} and accepting the MW null hypothesis $\mathcal{H}_{0}^{F}$:~\texttt{NON\_CATCH\_IC} $>$ \texttt{CATCH\_IC}. 
This represents additional evidence that EH code is considerably less covered than non-EH code.

However, when we look at the coverage inside \texttt{try} and \texttt{finally} blocks and their counterparts (i.e., the coverage of instructions outside \texttt{try} and \texttt{finally} blocks) we realize they are similar. 
This perception is confirmed by the statistical tests when both the KS and MW null hypotheses ($\mathcal{H}_{0}^{E}$: \texttt{NON\_TRY\_IC} $=$ \texttt{TRY\_IC} and $\mathcal{H}_{0}^{G}$: \texttt{NON\_FINALLY\_IC} $=$ \texttt{FINALLY\_IC}) are accepted.
Since \texttt{try} and \texttt{finally} blocks are commonly executed when no exceptional behavior is exercised, this observation corroborates with previous findings that code that handle exceptions are not properly tested.

\begin{figure}
	\centering
	\includegraphics[scale=0.80]{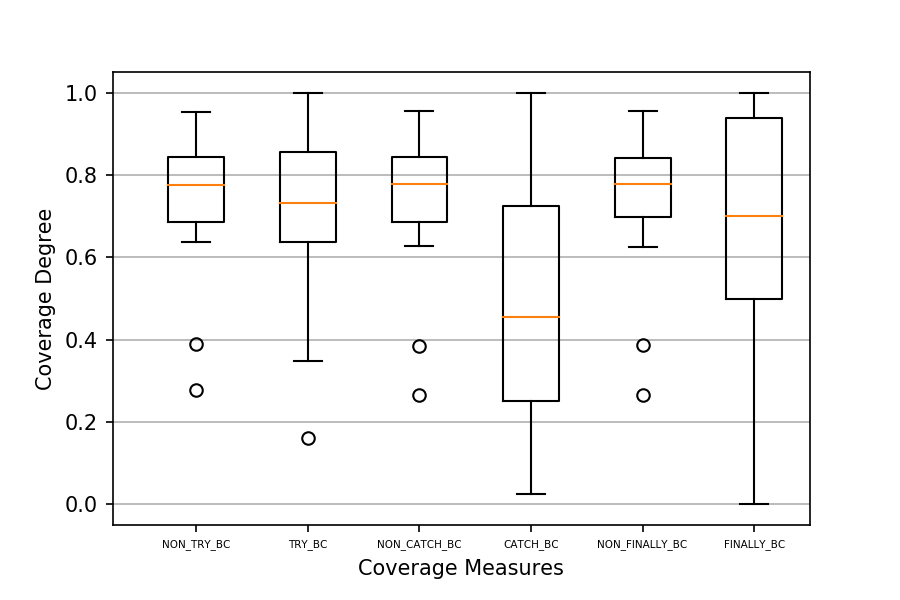}
	\caption{The EH and non-EH branch coverage boxplots of studied libraries.}
	\label{fig:EHBCTest}
\end{figure}

The boxplots of Fig.~\ref{fig:EHBCTest} shows the branch coverage distribution of EH and non-EH code. 
When comparing the boxplots of \texttt{CATCH\_BC} and \texttt{NON\_CATCH\_BC}, one must notice that branches inside \texttt{catch} blocks are less covered than branches outside \texttt{catch} blocks. 
Once more, the statistical test results confirm this perception by rejecting the KS null hypothesis $\mathcal{H}_{0}^{I}$: \texttt{NON\_CATCH\_BC} $=$ \texttt{CATCH\_BC} and accepting the MW null hypothesis $\mathcal{H}_{0}^{I}$: \texttt{NON\_CATCH\_BC} $>$ \texttt{CATCH\_BC}. 

However, when we compare the coverage of branches inside \texttt{try} and \texttt{finally} blocks with their counterparts (the coverage of branches outside \texttt{try} and \texttt{finally} blocks) we realize they are similar. 
This perception is also confirmed by the statistical tests when both the KS and MW null hypotheses ($\mathcal{H}_{0}^{H}$:~\texttt{NON\_TRY\_BC} $=$ \texttt{TRY\_BC} and $\mathcal{H}_{0}^{J}$: \texttt{NON\_FINALLY\_BC} $=$ \texttt{FINALLY\_BC}) are accepted.
Once again, all findings regarding branch coverage add to the observation that code which raises and handle exceptions are statistically less covered than regular code.

\subsection{RQ3. What is the effectiveness of EH testing in long-lived Java libraries?}\label{sec:RQ3Results}

\begin{center}
\cornersize{.2}
\Ovalbox{\begin{minipage}{.95\textwidth}
\textbf{Summary of RQ3:} The libraries under study present effective test suites for EH code, where 68\% of all the injected defects are identified. However, the libraries present difficulties in identifying defects in \texttt{finally} blocks.
\end{minipage}}
\end{center}

In this study, we employ mutation testing to assess the effectiveness of EH testing in the libraries under study, as detailed in Section~\ref{subsec:XaviEH}.
In Table~\ref{tab:ComputedMutationScores}, we present results of the mutation testing analysis we performed. 
For each mutation operator (see Section~\ref{sec:MutationOperators}), we show the number of mutants killed by the test suite, the number of mutants left alive, and the mutation score.

\begin{table}
\footnotesize
\renewcommand{\arraystretch}{1.165}
\caption{Details of the mutation testing analysis for each library. Column \texttt{L} indicates the number of live mutants, column \texttt{D} indicates the number of killed mutants, and \texttt{S} indicates the mutation score.}
\centering
\label{tab:ComputedMutationScores}
\begin{adjustbox}{angle=90}
\begin{tabular}{l|r|r|r|r|r|r|r|r|r|r|r|r|r|r|r|r|r|r|r|r|r|r|r|r}
    \toprule
    \multirow{2}{*}{\textbf{Library}} & 
    \multicolumn{3}{c|}{\texttt{CBI}} & 
    \multicolumn{3}{c|}{\texttt{CBD}} & 
    \multicolumn{3}{c|}{\texttt{CRE}} & 
    \multicolumn{3}{c|}{\texttt{FBD}} & 
    \multicolumn{3}{c|}{\texttt{PTL}} & 
    \multicolumn{3}{c|}{\texttt{CBR}} & 
    \multicolumn{3}{c|}{\texttt{TSD}} & 
    \multicolumn{3}{c}{\texttt{OVERALL}} \\
    
    & \multicolumn{1}{c|}{\texttt{L}} & \multicolumn{1}{c|}{\texttt{D}} &  \multicolumn{1}{c|}{\texttt{S}} & \multicolumn{1}{c|}{\texttt{L}} & \multicolumn{1}{c|}{\texttt{D}} & \multicolumn{1}{c|}{\texttt{S}} & \multicolumn{1}{c|}{\texttt{L}} & \multicolumn{1}{c|}{\texttt{D}} & \multicolumn{1}{c|}{\texttt{S}} & \multicolumn{1}{c|}{\texttt{L}} & \multicolumn{1}{c|}{\texttt{D}} & \multicolumn{1}{c|}{\texttt{S}} & \multicolumn{1}{c|}{\texttt{L}} & \multicolumn{1}{c|}{\texttt{D}} & \multicolumn{1}{c|}{\texttt{S}} & \multicolumn{1}{c|}{\texttt{L}} & \multicolumn{1}{c|}{\texttt{D}} & \multicolumn{1}{c|}{\texttt{S}} & \multicolumn{1}{c|}{\texttt{L}} & \multicolumn{1}{c|}{\texttt{D}} & \multicolumn{1}{c|}{\texttt{S}} & \multicolumn{1}{c|}{\texttt{L}} & \multicolumn{1}{c|}{\texttt{D}} & \multicolumn{1}{c}{\texttt{S}}\\
    
    \midrule
    \textbf{BCEL} & 1 & 6 & 0.86 & 23 & 113 & 0.83 & 23 & 113 & 0.83 & 4 & 1 & 0.20 & 0 & 3 & 1.00 & 1 & 11 & 0.92 & 110 & 296 & 0.73 & 162 & 543 & 0.77\\
    \textbf{BeanUtils} & 1 & 9 & 0.90 & 42 & 84 & 0.67 & 24 & 102 & 0.81 & 0 & 0 & 0.00 & 0 & 8 & 1.00 & 0 & 13 & 1.00 & 185 & 179 & 0.49 & 252 & 395 & 0.61\\
    \textbf{CLI} & 0 & 0 & 0.00 & 1 & 10 & 0.91 & 1 & 10 & 0.91 & 0 & 1 & 1.00 & 0 & 0 & 0.00 & 0 & 0 & 0.00 & 1 & 28 & 0.97 & 3 & 49 & 0.94\\
    \textbf{Codec} & 0 & 0 & 0.00 & 3 & 18 & 0.86 & 3 & 18 & 0.86 & 8 & 0 & 0.00 & 0 & 0 & 0.00 & 0 & 0 & 0.00 & 42 & 55 & 0.57 & 56 & 91 & 0.62\\
    \textbf{Collections} & 0 & 0 & 0.00 & 4 & 24 & 0.86 & 4 & 24 & 0.86 & 1 & 0 & 0.00 & 0 & 1 & 1.00 & 0 & 0 & 0.00 & 159 & 566 & 0.78 & 168 & 615 & 0.79\\
    \textbf{Compress} & 0 & 5 & 1.00 & 13 & 057 & 0.81 & 14 & 56 & 0.80 & 36 & 3 & 0.08 & 0 & 3 & 1.00 & 9 & 0 & 0.00 & 216 & 209 & 0.49 & 288 & 333 & 0.54\\
    \textbf{Configuration} & 1 & 6 & 0.86 & 3 & 125 & 0.98 & 3 & 125 & 0.98 & 2 & 93 & 0.98 & 0 & 4 & 1.00 & 0 & 4 & 1.00 & 21 & 264 & 0.93 & 30 & 621 & 0.95\\
    \textbf{DBCP} & 3 & 7 & 0.70 & 617 & 163 & 0.21 & 611 & 169 & 0.22 & 6 & 17 & 0.74 & 0 & 8 & 1.00 & 0 & 7 & 1.00 & 79 & 200 & 0.72 & 1316 & 571 & 0.30\\
    \textbf{DbUtils} & 0 & 0 & 0.00 & 11 & 20 & 0.65 & 11 & 20 & 0.65 & 3 & 17 & 0.85 & 0 & 1 & 1.00 & 0 & 0 & 0.00 & 30 & 16 & 0.35 & 55 & 74 & 0.57\\
    \textbf{Digester} & 4 & 1 & 0.20 & 22 & 41 & 0.65 & 22 & 41 & 0.65 & 5 & 2 & 0.29 & 0 & 3 & 1.00 & 1 & 3 & 0.75 & 79 & 31 & 0.28 & 133 & 122 & 0.48\\
    \textbf{Email} & 0 & 3 & 1.00 & 2 & 027 & 0.93 & 3 & 26 & 0.90 & 8 & 1 & 0.11 & 0 & 1 & 1.00 & 0 & 3 & 1.00 & 2 & 72 & 0.97 & 15 & 133 & 0.90\\
    \textbf{Exec} & 0 & 2 & 1.00 & 6 & 15 & 0.71 & 4 & 17 & 0.81 & 3 & 2 & 0.40 & 0 & 0 & 0.00 & 0 & 1 & 1.00 & 0 & 29 & 1.00 & 13 & 66 & 0.84\\
    \textbf{FileUpload} & 0 & 0 & 0.00 & 7 & 15 & 0.68 & 6 & 16 & 0.73 & 5 & 1 & 0.17 & 0 & 0 & 0.00 & 0 & 0 & 0.00 & 26 & 24 & 0.48 & 44 & 56 & 0.56\\
    \textbf{Functor} & 0 & 0 & 0.00 & 0 & 0 & 0.00 & 0 & 0 & 0.00 & 0 & 0 & 0.00 & 0 & 0 & 0.00 & 0 & 0 & 0.00 & 90 & 25 & 0.22 & 90 & 25 & 0.22\\
    \textbf{IO} & 0 & 2 & 1.00 & 40 & 38 & 0.49 & 46 & 32 & 0.41 & 2 & 6 & 0.75 & 0 & 1 & 1.00 & 0 & 2 & 1.00 & 37 & 255 & 0.87 & 125 & 336 & 0.73\\
    \textbf{Lang} & 0 & 2 & 1.00 & 13 & 058 & 0.82 & 15 & 56 & 0.79 & 1 & 4 & 0.80 & 0 & 3 & 1.00 & 0 & 1 & 1.00 & 88 & 292 & 0.77 & 117 & 416 & 0.78\\
    \textbf{Math} & 6 & 6 & 0.50 & 21 & 094 & 0.82 & 20 & 095 & 0.83 & 4 & 0 & 0.00 & 0 & 5 & 1.00 & 4 & 5 & 0.56 & 414 & 1080 & 0.72 & 469 & 1285 & 0.73\\
    \textbf{Net} & 1 & 5 & 0.83 & 31 & 129 & 0.81 & 26 & 134 & 0.84 & 9 & 15 & 0.63 & 1 & 5 & 0.83 & 0 & 3 & 1.00 & 47 & 112 & 0.70 & 115 & 403 & 0.78\\
    \textbf{Pool} & 1 & 3 & 0.75 & 8 & 057 & 0.88 & 8 & 057 & 0.88 & 46 & 33 & 0.42 & 0 & 7 & 1.00 & 0 & 1 & 1.00 & 15 & 64 & 0.81 & 78 & 222 & 0.74\\
    \textbf{Proxy} & 0 & 0 & 0.00 & 0 & 23 & 1.00 & 1 & 22 & 0.96 & 0 & 0 & 0.00 & 0 & 0 & 0.00 & 0 & 0 & 0.00 & 0 & 36 & 1.00 & 1 & 81 & 0.99\\
    \textbf{Validator} & 0 & 1 & 1.00 & 10 & 30 & 0.75 & 11 & 29 & 0.73 & 1 & 0 & 0.00 & 0 & 3 & 1.00 & 0 & 2 & 1.00 & 18 & 50 & 0.74 & 40 & 115 & 0.74\\
    \midrule
    \textbf{Gson} & 0 & 3 & 1.00 & 6 & 49 & 0.89 & 5 & 50 & 0.91 & 1 & 4 & 0.80 & 0 & 1 & 1.00 & 0 & 4 & 1.00 & 26 & 196 & 0.88 & 38 & 307 & 0.89\\
    \textbf{Hamcrest} & 0 & 0 & 0.00 & 1 & 11 & 0.92 & 1 & 11 & 0.92 & 0 & 0 & 0.00 & 0 & 0 & 0.00 & 0 & 0 & 0.00 & 7 & 12 & 0.63 & 9 & 34 & 0.79\\
    \textbf{Jsoup} & 0 & 0 & 0.00 & 2 & 30 & 0.94 & 2 & 30 & 0.94 & 1 & 2 & 0.67 & 0 & 3 & 1.00 & 0 & 0 & 0.00 & 11 & 35 & 0.76 & 16 & 100 & 0.86\\
    \textbf{JUnit} & 0 & 5 & 1.00 & 7 & 083 & 0.92 & 9 & 81 & 0.90 & 6 & 11 & 0.65 & 0 & 2 & 1.00 & 0 & 3 & 1.00 & 20 & 81 & 0.80 & 42 & 266 & 0.86\\
    \textbf{Mockito} & 1 & 10 & 0.91 & 4 & 076 & 0.95 & 7 & 73 & 0.91 & 7 & 13 & 0.65 & 0 & 0 & 0.00 & 0 & 0 & 0.00 & 25 & 211 & 0.89 & 44 & 383 & 0.90\\
    \textbf{X-Stream} & 1 & 2 & 0.67 & 36 & 196 & 0.84 & 32 & 200 & 0.86 & 8 & 11 & 0.58 & 2 & 19 & 0.90 & 1 & 1 & 0.50 & 160 & 301 & 0.65 & 240 & 730 & 0.75\\
    \bottomrule
    \end{tabular}
   \end{adjustbox}
\end{table}

\begin{figure}
	\centering
	\includegraphics[scale=0.80]{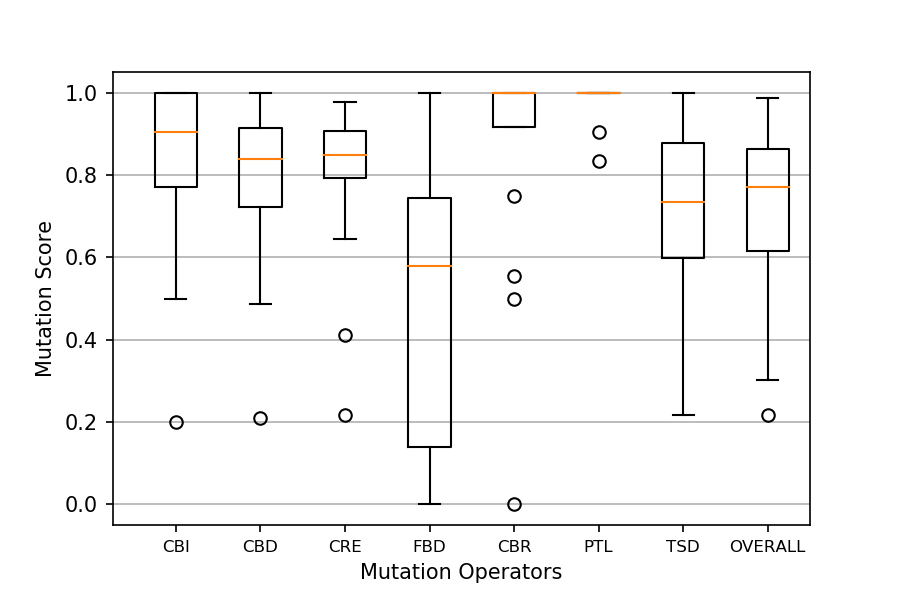}
	\caption{Mutation scores distribution bloxpots.}
	\label{fig:MutationScoreBloxpot}
\end{figure}

Furthermore, in Fig.~\ref{fig:MutationScoreBloxpot}, we present boxplots showing the mutation score distribution for all libraries and each mutation operator. 
Note that we computed the distributions considering only the mutation scores of libraries in which we were able to generate at least one mutant using the mutation operator associated with the boxplot. 
In this study, we generated a total of $12,331$ software mutants as follows: $98$ (\texttt{CBI}), $2,519$ (\texttt{CBD}), $2,519$ (\texttt{CRE}), $404$ (\texttt{FBD}), $84$ (\texttt{PTL}), $80$ (\texttt{CBR}), and $6,627$ (\texttt{TSD}). Considering all $12,331$ created mutants, we computed a global mutation score of $0.68$, which means that 68\% of all mutants were killed. Since we do not eliminate the equivalent mutants, this $0.68$ can be seen as a lower boundary value for the mutation score.

When looking at both the table and figure, one must notice that the libraries under study achieved mean and median mutation scores above 70\% for all but one mutation operator (\texttt{FBD}), which is considerably high when compared with other studies in the literature~\citep{Reales2014,Rahul2014,Laura2014}.
When taking into account all mutation operators, the median mutation score achieved is 78\%.
This indicates that the test suites in the studied libraries managed to detect a median of 78\% of artificially injected defects.
Operators such as \texttt{CBR} and \texttt{PTL}, for example, present median mutation scores of 100\%, indicating that the test suites of most of the libraries under study identified all bugs related to wrongly declared exceptions in \texttt{catch} blocks and wrongly placed instructions in \texttt{try} blocks.

Nevertheless, this is not the case for all mutation operators.
We observe a median mutation score of 59\% for the \texttt{FBD} operator, reaching even 0\% for some libraries.
This indicates that the libraries under study struggle in identifying defects in \texttt{finally} blocks.
This is an interesting observation because we showed in the previous research question that \texttt{finally} blocks are highly covered.
Hence, although being able to exercise EH code in \texttt{finally} blocks, the test suites have difficulties in actually identifying defects in them.

It is important to notice that not all mutation operators generated a similar number of mutants.
On the contrary, there are large differences between operators, such as \texttt{TSD} generating a total of 6,627 mutants for all libraries and \texttt{CBR} generating only 80 mutants overall.
This is due to how each operator generates mutants (see Section~\ref{sec:MutationOperators}).
While \texttt{TSD} simply deletes a \texttt{throw} statement, \texttt{CBR} identifies a \texttt{catch} block and searches the exception hierarchy to replace the exception for a derived type.
Hence, it is expected that there will be much more \texttt{throw} statements to be deleted throughout all libraries than derived exceptions in \texttt{catch} blocks to be replaced.
However, there seems to be no relationship between the number of mutants generated to the mutation score achieved.
For example, two operators with a small number of generated mutants, such as \texttt{FBD} and \texttt{PTL}, represent the operators with smallest and highest median of mutation score, respectively.
The relationship between number of mutants and the respective effectiveness of the test suite for this type of defects is still open for investigation.

\subsection{RQ4. To what extent are there EH bugs that are statistically harder to detect by test suites of long-lived Java libraries?}
\label{sec:RQ4Results}

\begin{center}
\cornersize{.2}
\Ovalbox{\begin{minipage}{.95\textwidth}
\textbf{Summary of RQ4:} There are EH bugs that are statistically harder to detect than others. In specfic, EH bugs of types \texttt{FBD}, \texttt{TSD}, \texttt{CRE}, and \texttt{CBD} are more difficult to detect than EH bugs of type \texttt{PTL}.
\end{minipage}}
\end{center}

To properly answer this research question, we employed a statistical test to verify whether there is any significant difference between the effectiveness of the studied libraries' test suites in detecting different types of artificially injected EH bugs (i.e., EH mutants). 
We used the Friedman test~\citep{Friedman1940}, which aims at quantifying the consistency of the results obtained by a test suite when applied over several types of EH bugs.
We applied each of the 7 mutation operators to each library, collecting the mutation score for each case. 
Next, we rank the 7 operators for each library according to their mutation score (the highest mutation score getting rank 1, the second-highest rank 2 and so on).
We leverage the Friedman test to check whether the mutation score for any of the 7 mutation operators ranks consistently higher or lower than the others.
In the current setting, the null hypothesis states that there is no statistical difference in detecting different types of EH bugs. 
If the Friedman null hypothesis is rejected, a post-hoc test must be applied to identify what type of EH bug is significantly easier/harder to detect than others. 
For this purpose, we adopt the post-hoc Nemenyi test~\citep{Demsar2006}. 


To ensure that the Friedman's test will yield significant results, the data points cannot present missing values.
Since XaviEH could not generate mutants for a few operators in some libraries (see Table~\ref{tab:ComputedMutationScores}), we selected for this analysis only the test suites of libraries that XaviEH could generate mutants for all mutation operators, which represents a total of $15$ studied libraries. 
Despite losing data points for this analysis, we can still observe statistically significant results because Friedman's test guidelines state that p-values are reliable for more than 6 measurements\footnote{\url{https://docs.scipy.org/doc/scipy/reference/generated/scipy.stats.friedmanchisquare.html}} (libraries' test suites in our study).

For each libraries' test suite, we used the set of all $7$ EH mutation operators and ranked them according to their mutation scores. 
Consider the \texttt{BCEL} library, for example.
The \texttt{PTL} operator presented the highest mutation score, which indicates that this was the easiest type of EH bug to detect in this library, yielding a rank 1.
Similarly, the \texttt{FBD} operator received the rank 7 because it presented the lowest mutation score for all operators in the \texttt{BCEL} library.
Next, we averaged the rankings for all mutation operators and produced the final average ranking. 
We present all computed rankings in Table~\ref{tab:AVGRankMutationScores}. 

\begin{table}
\renewcommand{\arraystretch}{1.2}
\caption{The ranks and average rank of mutation scores.}
\centering
\label{tab:AVGRankMutationScores}
\footnotesize
\begin{tabular}{l|r|r|r|r|r|r|r}
    \toprule
    \multicolumn{1}{c|}{\textbf{Library}} &  \multicolumn{1}{c|}{\texttt{CBI}} & \multicolumn{1}{c|}{\texttt{CBD}} & \multicolumn{1}{c|}{\texttt{CRE}} & \multicolumn{1}{c|}{\texttt{FBD}} & \multicolumn{1}{c|}{\texttt{PTL}} & \multicolumn{1}{c|}{\texttt{CBR}} & \multicolumn{1}{c}{\texttt{TSD}}\\
    \midrule
    
    \textbf{BCEL}          & $3.0$ & $4.5$ & $4.5$ & $7.0$ & $1.0$ & $2.0$ & $6.0$\\
    
    \textbf{Compress}      & $1.5$ & $3.0$ & $4.0$ & $6.0$ & $1.5$ & $7.0$ & $5.0$\\
    
    \textbf{Configuration} & $7.0$ & $4.5$ & $4.5$ & $3.0$ & $1.5$ & $1.5$ & $6.0$\\

    \textbf{DBCP}          & $5.0$ & $7.0$ & $6.0$ & $3.0$ & $1.5$ & $1.5$ & $4.0$\\

    \textbf{Digester}      & $7.0$ & $3.5$ & $3.5$ & $5.0$ & $1.0$ & $2.0$ & $6.0$\\

    \textbf{Email}         & $2.0$ & $5.0$ & $6.0$ & $7.0$ & $2.0$ & $2.0$ & $4.0$\\

    \textbf{IO}            & $2.0$ & $6.0$ & $7.0$ & $5.0$ & $2.0$ & $2.0$ & $4.0$\\

    \textbf{Lang}          & $2.0$ & $4.0$ & $6.0$ & $5.0$ & $2.0$ & $2.0$ & $7.0$\\

    \textbf{Math}          & $6.0$ & $3.0$ & $2.0$ & $7.0$ & $1.0$ & $5.0$ & $4.0$\\

    \textbf{Net}           & $3.5$ & $5.0$ & $2.0$ & $7.0$ & $3.5$ & $1.0$ & $6.0$\\

    \textbf{Pool}          & $6.0$ & $3.5$ & $3.5$ & $7.0$ & $1.5$ & $1.5$ & $5.0$\\

    \textbf{Validator}     & $2.0$ & $4.0$ & $6.0$ & $7.0$ & $2.0$ & $2.0$ & $5.0$\\
    \midrule
    
    \textbf{Gson}          & $2.0$ & $5.0$ & $4.0$ & $7.0$ & $2.0$ & $2.0$ & $6.0$\\

    \textbf{JUnit}         & $2.5$ & $4.0$ & $5.0$ & $7.0$ & $2.0$ & $2.0$ & $6.0$\\

    \textbf{X-Stream}      & $4.0$ & $3.0$ & $2.0$ & $6.0$ & $1.0$ & $7.0$ & $5.0$\\
    \midrule
    \textbf{Average Rank}  & $\mathbf{3.7}$ & $\mathbf{4.3}$ & $\mathbf{4.4}$ & $\mathbf{5.9}$ & $\mathbf{1.7}$ & $\mathbf{2.7}$ & $\mathbf{5.3}$\\
    \bottomrule
\end{tabular}
\end{table}

According to the Friedman test, the average ranking difference is significant with p-value $ = 1.2 \times 10^{-4}$. 
Hence, we attest that there exists types of EH bugs that are statistically harder to identify by the test-suites under study.
Next, we employed Nemenyi's post-hoc test, which showed a Critical Difference of \texttt{CD} = $2.32$. 
In this context, the performance of two mutation operators is said to be significantly different if their average ranking differ by at least the \texttt{CD} level.
The \texttt{CD} metric is computed using Equation (\ref{eq:cd}), where $k$ is the number of mutation operators, $N$ is the number of test suites (libraries), and $q_{\alpha}$ is a pre-calculated critical value that one must pick up from a reference table by observing the value of $k$ and the confidence interval ($\alpha$). 
Thus, for our study $k = 7$, $N = 15$, $\alpha = 0.05$, and $q_{0.05} =  2.948$.

\begin{equation}\label{eq:cd}
    \mathtt{CD} = q_{\alpha}\sqrt{\frac{k (k + 1)}{6N}}
\end{equation}

Based on these results, we can conclude that the \texttt{FBD} and \texttt{TSD} mutation operators generate EH bugs that are statistically more difficult to detect than EH bugs generated by the \texttt{PTL} and \texttt{CBR} mutation operators. 
Additionally, the \texttt{CRE} and \texttt{CBD} operators generate EH bugs that are significantly harder to detect than EH bugs generated by the \texttt{PTL} operator.
Even though we observe differences in the ranking between \texttt{FBD}, \texttt{TSD}, \texttt{CRE}, and \texttt{CBD}, we cannot ascertain significant statistical difference between them.
This indicates that, according to our empirical study, these are equally the most difficult types of EH bugs to detect.



\section{Discussion}\label{sec:discussion}

In this section, we sum up the most important findings of our empirical study and discuss their implications.
Finally, we briefly discuss how XaviEH could be used in practice.

\subsection{On the Adequacy of EH Testing}

In RQ1-2, we present empirical and statistical evidence that EH code is less covered than regular code in the libraries under study.
Moreover, we show that within coverage of EH code, instructions and branches inside \texttt{catch} blocks and \texttt{throw} instructions are statistically less covered than instructions and branches in \texttt{try} and \texttt{finally} blocks.
This indicates that not only these test suites do not properly cover EH code (statements in \texttt{catch} blocks) but also that these suites are not able to exercise the code parts responsible for raising exceptions (\texttt{throw} statements).
We followed on this insight by computing two sets of Spearman's correlations considering all the libraries under study.

The first correlation was computed between \texttt{throw} instruction coverage (\texttt{THROW\_IC}) and \texttt{catch} blocks' instruction coverage (\texttt{CATCH\_IC}), and the second between \texttt{throw} instruction coverage (\texttt{THROW\_IC}) and \texttt{catch} blocks branch (\texttt{CATCH\_BC}) coverage. 
The results show a strong correlation in both cases with $\rho = 0.582664$ (\texttt{THROW\_IC} and \texttt{CATCH\_IC}) and $\rho = 0.674882$ (\texttt{THROW\_IC} and \texttt{CATCH\_BC}). 
The correlation results confirm that these coverage values are strongly connected, as empirically observed.
We suggest two possible scenarios that may explain such correlations.
In the first scenario, the \texttt{catch} blocks not being covered are the ones responsible for catching the exceptions not being raised.
Differently, in the second scenario, the two elements (\texttt{throw} statements and \texttt{catch} blocks) are not connected, which may include JRE and other external exceptions not being caught, for instance.
Further studies are needed to thoroughly understand this phenomenon.

Our study shows that developers need better support in designing test cases that exercise exceptional behaviors.
In addition to creating guidelines, this may be accomplished through search-based testing~\citep{mcminn2004search}, where optimization algorithms and metaheuristics are used to generate test cases according to a certain objective function automatically.
In this case, one could set the coverage of \texttt{throw} instructions and branches and instructions inside \texttt{catch} blocks as a goal.
To the best of our knowledge, there is existing work in this direction~\citep{romano2011approach}.

\subsection{On the Effectiveness of EH Testing}
\label{sec:OnEffectiveness}

RQ3-4 shows that despite not properly covering EH code, the test suites of the libraries under study are surprisingly effective in identifying artificially injected faults (EH mutants).
Most of the libraries presented mutation scores of more than $68\%$ for most mutation operators.
However, this was not the case for all operators.
In fact, we showed that there do exist statistically harder types of EH bugs to identify.
These are commonly related to mutations in \texttt{throw} statements and \texttt{catch} and \texttt{finally} blocks.
This is an interesting finding that corroborate with what we have previously discussed.
The code in EH mechanism that actually raises (\texttt{throw} statements) and handles exceptions (statements in \texttt{catch} blocks) seems to be the most fragile, in which it is less covered and more difficult to identify faults.

\subsection{On Qualitatively Assessing our Results}

Despite our paper having the main goal of quantitatively assessing EH testing practices in real-world libraries, we found a few interesting cases among the data in which a closer inspection may yield interesting insights.

First, we noticed that the coverage results for the different libraries presented a wide variation.
For this analysis, we take the value of \texttt{TRY\_CATCH\_BC} as our metric of comparison.
We chose this metric because it is a good representative of EH testing adequacy, as previously discussed in our results sections.
Based on this metric, \texttt{CLI} presented the best coverage results with 73\% of its \texttt{catch} blocks being covered.
As its counterpart, \texttt{BCEL} only covers 2\% of its \texttt{catch} blocks.
This may be explained by the size of each library.
While \texttt{BCEL} is composed of 344 classes and contains 143 \texttt{catch} blocks, \texttt{CLI} is composed of 21 classes and contains only 11 \texttt{catch} blocks.
However, this behavior is not repeated through the whole dataset.
\texttt{Math}, for instance, is the largest library with 740 classes, and it covers 47\% of its 180 \texttt{catch} blocks.
We suggest that further studies are needed to fully understand this phenomenon.

In our study, we collected data from Apache and non-Apache libraries. Although the number of libraries in both subsets does not allow for statistical comparisons, we can still observe a few interesting details. All of the non-Apache libraries achieve less than 50\% \texttt{catch} blocks coverage, as according to \texttt{TRY\_CATCH\_BC}. Differently, 5 out of 21 Apache libraries cover more than half of its \texttt{catch} blocks. This observation is especially interesting because 3 of the non-Apache libraries are testing-related frameworks (\texttt{Hamcrest}, \texttt{JUnit} and \texttt{Mockito}). This attests to the well-known quality control of the Apache ecosystem.

\subsection{On the Usefulness of XaviEH}

Our empirical study was powered by XaviEH, a tool that automatically generates a complete analysis and report of EH coverage and mutation testing for a certain Java system.
XaviEH can be easily employed by developers as an EH testing diagnostics tool.
Based on XaviEH outputs, developers can plan and improve their test suites regarding EH code.

Furthermore, given its full automated features, XaviEH could be also accommodated in continuous integration pipelines.
In this context, developers would receive EH testing reports in each commit, which could create and foster a culture of continuous improvement of EH testing practices.
In addition, XaviEH's outputs could be employed as metrics and proxies of testing effectiveness, as well as goals to be achieved by the development team.

\section{Threats to Validity}\label{sec:threats}

The threats to the validity of our investigation are discussed using the four threats classification (conclusion, construct, internal, and external validity) presented by~\cite{Wohlin2012}.

\subsection{Conclusion Validity}

Threats to the conclusion validity are concerned with issues that affect the ability to draw correct conclusions regarding the treatment and the outcome of an experiment. 
To deal with this threat, we carefully chose proper statistical tests (KS, MW, Friedman, and Nemenyi tests) that have been investigated and validated in previous studies \citep{kumar2011new}, \citep{ji2009new}. 
We also selected correlation measures (Spearman’s rank-order correlation coefficient) to investigate the relationship between different aspects of EH testing (by means of code coverage and mutation analysis) and its effectiveness. 
Additionally, we have observed the assumptions (e.g., samples distribution, dependence, and size) of all statistical tests we used, trying to avoid wrong conclusions. 
Finally, regarding the limited set of long-lived Java libraries, we collected them from open-source communities following a carefully defined set of criteria to ensure the disposal of other libraries that were not aligned with the study.

\subsection{Internal Validity}

Threats to internal validity are influences that can affect the independent variable with respect to causality, without the researcher’s knowledge. 
Thus they threat the conclusion about a possible causal relationship between treatment and outcome. 
Even not being interested in drawing causal relationships in our study, we have identified that some independent variables not known by us have some influence on the relationship between the EH code coverage and the mutation scores distribution in the studied libraries. 

\subsection{Construct Validity}

Construct validity concerns generalizing the result of the study to the concept or theory behind the study. 
To avoid inconsistencies in the interpretation of the results and research question, a peer debriefing approach was adopted for both research design validation and document review. 
Additionally, we developed a tool, XaviEH, in order to automate most of the study's parts, with an aim to avoid or alleviate the occurrence of human-made mistakes (or bias) during the execution of our experiments.

\subsection{External Validity}

Threats to external validity are conditions that limit our ability to generalize the results of our study to industrial practice. 
The main threats to this validity are related to the domain and sample size (i.e., the 27 libraries) we used in this study. 
Concerning the sample domain, we try to deal with this threat by arguing that the library domain is an interesting one that presents several and different usage scenarios, which is quite interesting from a testing evaluation point of view. 
Additionally, concerning the sample size, we dealt with this threat by using diversity and longevity criteria. 
We chose libraries from Apache and picked up other well-know libraries developed by other development teams to get more diversity in terms of team knowledge, skills, and coding practices. 
Finally, we chose libraries that are long-lived as a way to guarantee a degree of maturity and stability.

\section{Related Work}\label{sec:relatedwork}

In this section, we present the related work that, in some way, are related to our study.

\subsection{Exception Handling and Software Bugs}\label{subsection:ehsb}

Previous work has investigated and provided evidence on the positive correlation between exception handling code and software defect proneness~\citep{Marinescu2011, Marinescu2013}. This correlation emerges from sub-optimal exception handling practices (i.e., anti-patterns and flow characteristics) current adopted by software developers~\citep{Sawadpong2016,dePadua2018}. Additionally, the exception handling is usually neglected by developers (mainly by novices ones) and is considered as one of the least understood, documented, and tested part of a software system~\citep{Shah2010,Zhang2014,chang2016review,Oliveira2018}.

The studies conducted by \cite{Barbosa2014} and \cite{Ebert2015} gather evidence that erroneous or improper usage of exception handling can lead to a series of fault patterns, named ``exception handling bugs''. This kind of faults refer to bugs in which the primary source is related to (i) the exception definition, throwing, propagation, handling or documentation; (ii) the implementation of cleanup actions; and (iii) wrong throwing or handling (i.e., when the exception should be thrown or handled and it is not). \cite{Barbosa2014} categorizes 10 causes of exception handling bugs, analyzing two open-source projects, Hadoop and Apache Tomcat. \cite{Ebert2015} extends \cite{Barbosa2014} study, presenting a comprehensive classification of exception handling bugs based on a survey of 154 developers and the analysis of 220 exception handling errors reported from two open-source projects, Apache Tomcat and Eclipse IDE. \cite{Kechagia2014} studied undocumented runtime exceptions thrown by the Android platform and third-party libraries. They mined 4,900 different stack traces from 1,800 apps looking for undocumented API methods with undocumented exceptions participating in the crashes. They found that 10\% of crashes might have been avoided if the correspondent runtime exceptions had been properly documented. 

~\cite{dePadua2017, dePadua2017b, dePadua2018} conducted a series of studies concerning exception handling and software quality. In the first study, they conducted an investigation on the prevalence of exception handling anti-patterns across 16 open-source projects (Java and C\#). They claim that the misuse of exception handling can cause catastrophic software failures, including application crashes. They found that all 19 exception handling anti-patterns taken into account in the study are broadly present in all subject projects. However, only 5 of them (unhandled exception, generic catch, unreachable handler, over-catch, and destructive wrapping) are prevalent. Next, \cite{dePadua2017b} conducted a study revisiting the exception handling practices by analyzing the flow of exceptions from the source of exceptions until its handling blocks in 16 open-source projects (Java and C\#). Once researchers understood that exception handling practices might lead to software failures, their identification highlight the opportunities of leveraging automated software analysis to assist in exception handling practices. 

~\cite{dePadua2018} focuses on understanding the relationship between exception handling practices and post-release defects. They investigated the relationship between post-release defect proneness and: (i) exception flow characteristics; and (ii) 17 exception handling anti-patterns. Their finds suggest that development teams should find a way to improve their exception handling practices and avoid the anti-patterns (e.g., dummy handler, generic catch, ignoring interrupted exception, and log and throw), that are found to have a relationship with post-release defects.

~\cite{Coelho2017} mined 6,000 stack traces from over 600 open-source projects issues on GitHub and Google Code searching for bug hazards regarding exception handling. Additionally, they surveyed 71 developers involved in at least one of the projects analyzed. As a result, they found four bug hazards that may cause bugs in Android applications: (i) cross-type exception wrapping; (ii) undocumented unchecked exceptions raised by the Android platform and third-party libraries; (iii) undocumented check exceptions signaled by native C code; and (iv) programming mistakes made by developers. The survey's results corroborate the stack trace findings, indicating that developers are unaware of frequently occurring undocumented exception handling behavior.

Similar to the mentioned studies, our study investigates EH testing practices in 27 long-lived Java libraries with more than 11 years of active development. We generated a total of $12,331$ software mutants and observed that the systems present effective test suites for EH code, where more than 68\% of the defects were identified. However, the libraries present difficulties in identifying defects in \texttt{finally} blocks.

\subsection{Exception Handling Testing}

\cite{ji2009new} proposes 5 types of exception handling code mutants: Catch Block Replacement (\texttt{CBR}), Catch Block Insertion (\texttt{CBI}), Catch Block Deletion (\texttt{CBD}), Placing Try Block Later (\texttt{PTL}), Catch and Rethrow Exception (\texttt{CRE}). \cite{kumar2011new} develops 5 types of mutants for exception handling code, namely: Catch Clauses Deletion (\texttt{CCD}), Throw Statement Deletion (\texttt{TSD}), Exception Name Change (\texttt{ENC}), Finally Clause Deletion (\texttt{FCD}) and Exception Handling Modification (\texttt{EHM}). These operators try to replace, insert, delete some \texttt{catch} blocks, add statements to re-throws a caught exception, and try to rearrange \texttt{try} blocks by including statements with some relevant references after the \texttt{catch} blocks. In our study, we employed 7 mutation operators, 5 (\texttt{CBR}, \texttt{CBI}, \texttt{CBD}, \texttt{PTL}, and \texttt{CRE})  from~\citep{ji2009new} and 2 (\texttt{FCD} we call \texttt{FBD} and \texttt{TSD}) from~\citep{kumar2011new}.

\cite{Zhang2014} presents an automated approach to support the detection of faults in exception handling code that deals with external resources. The study revealed that $22$\% of the confirmed and fixed bugs have to do with poor exceptional handling code, and half of those correspond to interactions with external resources. In our study, we identified as a result that despite presenting high coverage for instruction and branches in the overall source code and EH code, the tests are still mostly exercising non-exceptional flows within the programs, where the exception behaviors are not being tested.

\cite{goffi2016} presented a technique to automatically generate test oracles for exceptional behavior, called Toradocu. Toradocu uses natural language processing to automatically extract conditional expressions regarding exceptional behavior from Javadoc comments. An empirical evaluation shows that Toradocu improves the fault-finding effectiveness of EvoSuite and Randoop test suites by 8\% and 16\% respectively, and reduces EvoSuite’s false positives by 33\%.

\cite{Zhai2019} undertook a study of code coverage in popular Python projects: flask, matplotlib, pandas, scikit-learn and scrapy. In this study, the authors found that coverage depends on the control flow structure, with more deeply nested statements being significantly less likely to be covered. Other findings of the study were that the age of a line per se has a small (but statistically significant) positive effect on coverage. Finally, they found that the kind of statement (e.g., try, if, except, and raise) has a varying impact on coverage, with exception handling statements being covered much less often. The results suggest that developers in Python projects have difficulty writing test sets that cover deeply-nested and error-handling statements, and might need assistance covering such code. 

\cite{dalton2020ehtest} performed a study to understand how 417 open source Java projects are testing the exceptional behavior. They looked at test suites coded using JUnit and TestNG frameworks, and the AssertJ library. Overall, they count test methods that expect exceptions to be raised, which they called ``exceptional behavior testing''. They found that (i) 60.91\% of projects have at least one test method dedicated to testing the exceptional behavior; (ii) the number of test methods for exceptional behavior with respect to the total number of test methods lies between 0\% and 10\% in 76.02\% of projects; and (iii) 57.31\% of projects test only up to 10\% of the used exceptions in the system under test. They triangulate such results with a survey with 66 developers from the studied projects. The survey respondents confirm the findings and support the claim that developers often neglect exceptional behavior tests. However, different from our study, \cite{dalton2020ehtest} did not evaluate the adequacy and effectiveness of exceptional behavior testings itself (i.e., did not run the test methods that exercise the exceptional behaviors of the system under test). Instead, they performed a static analysis of the testing code to gather evidence on whether or not there are test methods intended to test the exceptional behavior.

Our study used long-lived Java libraries and our results shows that in spite of the low coverage for instruction and branches related to EH code, the unit tests of the studied libraries were able to detect a significant amount of artificially injected faults.

\subsection{Code Coverage}

\cite{gligoric2013} presented an extensive study that evaluates coverage criteria over non-adequate test suites. The authors analyzed a large set of plausible criteria, including statement and branch coverage, as well as stronger criteria used in recent studies. Two criteria performed best: branch coverage and intra-procedural acyclic path coverage. The study's results suggest that researchers should use branch coverage to compare suites whenever possible, but all evaluated criteria performed well to predict mutation scores.

\cite{Inozemtseva2014} conducted one of the first large studies that investigated the correlation between code coverage and test effectiveness. Their study took into account 31,000 test suites generated for five large Java systems. They measured code coverage (statement, branch, and modified condition) using these test suites and employed mutation testing to evaluate the effectiveness of such test suites in revealing the injected faults. They found that there is a low to moderate correlation between coverage and effectiveness when the number of test cases in the suite is controlled for this purpose. \cite{Kochhar2015} conducted a study seeking out to investigate the correlation between code coverage and its effectiveness in real bugs. The experiment was performed, taking into account 67 and 92 real bugs from Apache HTTPClient and Mozilla Rhino, respectively. They used a tool called Randoop, to generate random test suites with varying levels of coverage and run them to analyze the capability of these synthetic test suites in detecting the existing bugs in both systems. They found that there is a statistically significant correlation between code coverage and bug detection effectiveness. 

\cite{chekan2017} conducted a study using a robust experimental methodology that provided evidence to support the claim that strong mutation testing yields high fault revelation, while statement, branch and weak mutation testing enjoy no such fault revealing ability. The findings also revealed that only the highest levels of strong mutation coverage attainment have strong fault-revealing potential.

\cite{Kochhar2017} performed a large scale study concerning the correlation between real bugs and code coverage of exiting test suits. This study took into account 100 large open-source Java projects. They extracted real bugs recorded in the project's issue tracking system after the software release and analyzed the correlations between code coverage and these bugs. They found that the coverage of actual test suites has an insignificant correlation with the number of bugs that are found after the software release. 

\cite{Schwartz2018} argues that previous work provides mix results concerning the correlation between code coverage and test effectiveness (i.e., some studies provide evidence on a statistically significant correlation between these two factors, while others do not). Thus, they hypothesize that the fault type is one of the sources that may be leading to these mixed results. To investigate this hypothesis, they have studied 45 different types of faults and evaluated how effectively human-created test suites with high coverage percentages were able to detect each type of fault. The study was performed on 5 open-source projects (Commons Compress, Joda Time, Commons Lang, Commons Math, and JSQL Parser), which have at least 80\% statement coverage. The mutation testing technique was employed to seed 45 types of faults in the program's code to evaluate the effectiveness of the existing unit test suites in the detection of such fault types. Their findings showed that, with statistical significance, there were specific types of faults found less frequently than others. Additionally, based on their findings, they suggest developers should put more focus on improving test oracles strength along with code coverage to achieve higher levels of test effectiveness.

Our study analyzes the test coverage compared to the EH code. We developed a tool, called XaviEH, which employs both coverage and mutation analysis as proxies for the effectiveness of EH testing in a certain libraries. Our findings suggest that EH code is, in general, less covered than regular code (i.e., non-EH).

\section{Conclusion and Final Remarks}\label{sec:conclusion}

In this study, we empirically explored EH testing practices by analyzing in which degree the EH code is covered by unit-test suites of 27 long-lived java libraries and how effective these test suites are in detecting artificially injected EH faults. Our findings suggest that, indeed, EH code is, in general, less covered than non-EH code. Additionally,  we gather evidence indicating that the code within \texttt{catch} blocks and the \texttt{throw} statements have a low coverage degree. However, even being less covered, the mutation analysis shows that the test suites can detect most of the artificial EH faults.

To the best of our knowledge, this is the first study that empirically addresses this concern. Thus, the results achieved in this study can be seen as a starting point for further investigation regarding testing practices for EH code.

This study was deeply supported by the XaviEH tool. Without this level of automation, it would not be possible to manually extract and synthesize information regarding EH code coverage and EH mutation scores. Therefore, we freely turn it available to the community~\citep{replication_package}.
We include XaviEH's source code and usage instructions.

As future work, we are interested in (i) investigating the performance of the libraries test suites against real-world bugs; (ii) investigating the performance of the libraries test suites in a software evolution scenario; (iii) exploring libraries from different domains; and (iv) inspecting the test suites to identify and catalog what practices make a test suite better than others regarding EH testing.

\bibliographystyle{spbasic}      
\bibliography{references}

\end{document}